\documentclass[12pt,preprint]{aastex}

\usepackage{graphicx}
\usepackage{epstopdf}

\usepackage{natbib}
\newcommand{\coone}{$\rm ^{12}CO(1-0)$}
\newcommand{\cotwo}{$\rm ^{12}CO(2-1)$}
\newcommand{\cotre}{$\rm ^{13}CO(1-0)$}
\newcommand{\coeig}{$\rm C^{18}O(1-0)$}
\newcommand{\cotretwo}{$\rm ^{13}CO(2-1)$}

\newcommand{\hcn}{HCN(1-0)}
\newcommand{\hco}{HCO+(1-0)}

\newcommand{\kms}{\>{\rm km}\,{\rm s}^{-1}}

\newcommand{\as}{^{\prime\prime}}

\newcommand{\bdm}{\begin{displaymath}}
\newcommand{\edm}{\end{displaymath}}
\newcommand{\beq}{\begin{equation}}
\newcommand{\eeq}{\end{equation}}
\newcommand{\bit}{\begin{itemize}}
\newcommand{\eit}{\end{itemize}}
\newcommand{\ben}{\begin{enumerate}}
\newcommand{\een}{\end{enumerate}}
\newcommand{\bfi}{\begin{figure}[htb]}
\newcommand{\bpfi}{\begin{figure}[p]}

\shorttitle{Molecular Gas in M\,51}
\shortauthors{Schinnerer et al.}

\begin{document}

\title{Multi-Transition Study of M51's Molecular Gas Spiral Arms}


\author{E. Schinnerer}
\affil{MPI for Astronomy, K\"onigstuhl 17, 69117 Heidelberg, Germany}

\author{A. Wei\ss}
\affil{MPI for Radioastronomy, Auf dem H\"ugel 69, 53121 Bonn, Germany}

\author{S. Aalto}
\affil{Chalmers University of Technology, Department of Radio and 
Space Science, SE-412 96 G\"oteborg, Sweden}

\and

\author{N.Z. Scoville}
\affil{California Institute of Technology, MC 249-17, Pasadena, CA 91125, 
U.S.A.
}

\begin{abstract}
  Two selected regions in the molecular gas spiral arms in M\,51 were
  mapped with the Owens Valley Radio Observatory (OVRO)
  mm-interferometer in the \cotwo, \cotre, \coeig, HCN(1-0) and
  HCO+(1-0) emission lines. The CO data have been combined with the
  \coone\ data from \cite{aal99} covering the central 3.5\,kpc to
  study the physical properties of the molecular gas. All CO data
  cubes were short spacing corrected using IRAM 30m (\coone: NRO 45m)
  single dish data. A large velocity gradient (LVG) analysis finds
  that the giant molecular clouds (GMCs) are similar to Galactic GMCs
  when studied at 180\,pc (120\,pc) resolution with an average kinetic
  temperature of $\rm T_{kin} = 20(16)\,K$ and H$_2$ density of $\rm
  n(H_2) = 120(240)\,cm^{-3}$ when assuming virialized clouds (a
  constant velocity gradient $\rm \frac {d\it{v}}{d\it{r}}$). The
  associated conversion factor between $H_2$ mass and CO luminosity is
  close to the Galactic value for most regions analyzed. Our findings
  suggest that the GMC population in the spiral arms of M\,51 is
  similar to those of the Milky Way and therefore the strong star
  formation occurring in the spiral arms has no strong impact on the
  molecular gas in the spiral arms. Extinction inferred from the
  derived $H_2$ column density is very high ($A_V$ about 15 - 30 mag),
  about a factor of 5-10 higher than the average value derived toward
  HII regions.  Thus a significant fraction of the ongoing star
  formation could be hidden inside the dust lanes of the spiral arms.
  A comparison of MIPS 24$\mu$m and H$\alpha$ data, however, suggests
  that this is not the case and most of the GMCs studied here are not
  (yet) forming stars. We also present low (4.5'') resolution OVRO maps
  of the HCN(1-0) and HCO+(1-0) emission at the location of the
  brightest \coone\ peak.
\end{abstract}


\section{Introduction}

Since the first discovery of giant molecular clouds (GMCs) in our
Galaxy, a major aim has been to understand their range of physical
properties and their dependence on environment. Recently, \cite{bol08}
showed that the GMC properties are remarkably similar over a large
range of environments using a large set of interferometric
observations of local dwarf galaxies as well as M31 and M33. Most
studies of spiral galaxies outside the Local Group were undertaken
using single dish observations where the GMCs are no longer resolved.
Thus it is not obvious if the results of these studies can be directly
applied to the GMC population present in these galaxies. Detailed
analysis of the molecular gas at roughly GMC-scale resolution were
only obtained for the centers of a few nearby galaxies such as
IC\,342, Maffei\,2 and M\,82 \citep{mei01,mei08,wei01} all suggesting
that the properties of these GMCs differ from those found in galaxy
disks by exhibiting higher kinetic temperatures and lower conversion
factors likely due to the effect of the more vigorous star formation
present in the centers of these galaxies. Thus it would be
interesting to know whether GMCs residing in galactic disks with
enhanced star formation are similar to local GMCs or closer to those
found in the central regions of external galaxies.

The Whirlpool galaxy M\,51 is an ideal target being one of the closest
\citep[D=8.4\,Mpc with 1'' $\sim$ 40.7\,pc; ][]{fel97} almost face-on
grand-design spiral galaxies. A significant fraction of its molecular
gas is found in the spectacular spiral arms
\citep[e.g.][]{sco83,aal99,shet07,kod09}.  Therefore it offers an
unique opportunity to study the physical properties of the molecular
gas within the spiral arms. For example, \cite{kra05} combined single
dish data with observations from ISO to investigate the state of the
interstellar medium (ISM) in the center and selected arm regions
finding indications that the ISM in M\,51 might differ from that of
the Galaxy as most of the [CII] line emission in M\,51 can be
explained by arising from a photon-dominated region (PDR) unlike the
50\% assumed for the Galaxy. Due to M51's low inclination ($i \sim$
20$^o$) non-circular motions can be more easily separated from the
galactic rotation. The unusually large streaming motions
(60\,-\,150\,km\,$s^{-1}$) found in M\,51 imply a very strong density
wave whose general features are velocity discontinuities and streaming
across the spiral arms \citep[e.g.][]{rob87}.

Besides the large number of Giant Molecular Cloud Associations (GMAs)
\citep[up to 16 within one spiral arm,][]{aal99}, numerous star
clusters and HII regions reside in the spiral arms as is obvious in
high resolution HST imaging \citep[e.g.][]{sco01}. These populations have
been extensively studied using HST data \citep[e.g.][ and references
therein]{bas05,lee05,sche09}. Two interesting results in the context
of this paper are that younger clusters show higher extinction and are
located closer to spiral arms. 

The paper is organized as follows, the observations are presented in
\S \ref{sec:obs} while the data are described in \S \ref{sec:co}. A
non-LTE analysis of the spectral lines is given in \S \ref{sec:lvg}.
The relation between the GMC properties and star formation is
investigated in \S \ref{sec:sf}. A general discussion (\S
\ref{sec:dis}) and a summary of the findings (\S \ref{sec:sum}) are
provided at the end.

\section{Observations}
\label{sec:obs}

\subsection{Millimeter Interferometric Observations}

\begin{figure}
\centering
\includegraphics[height=20cm,angle=-90]{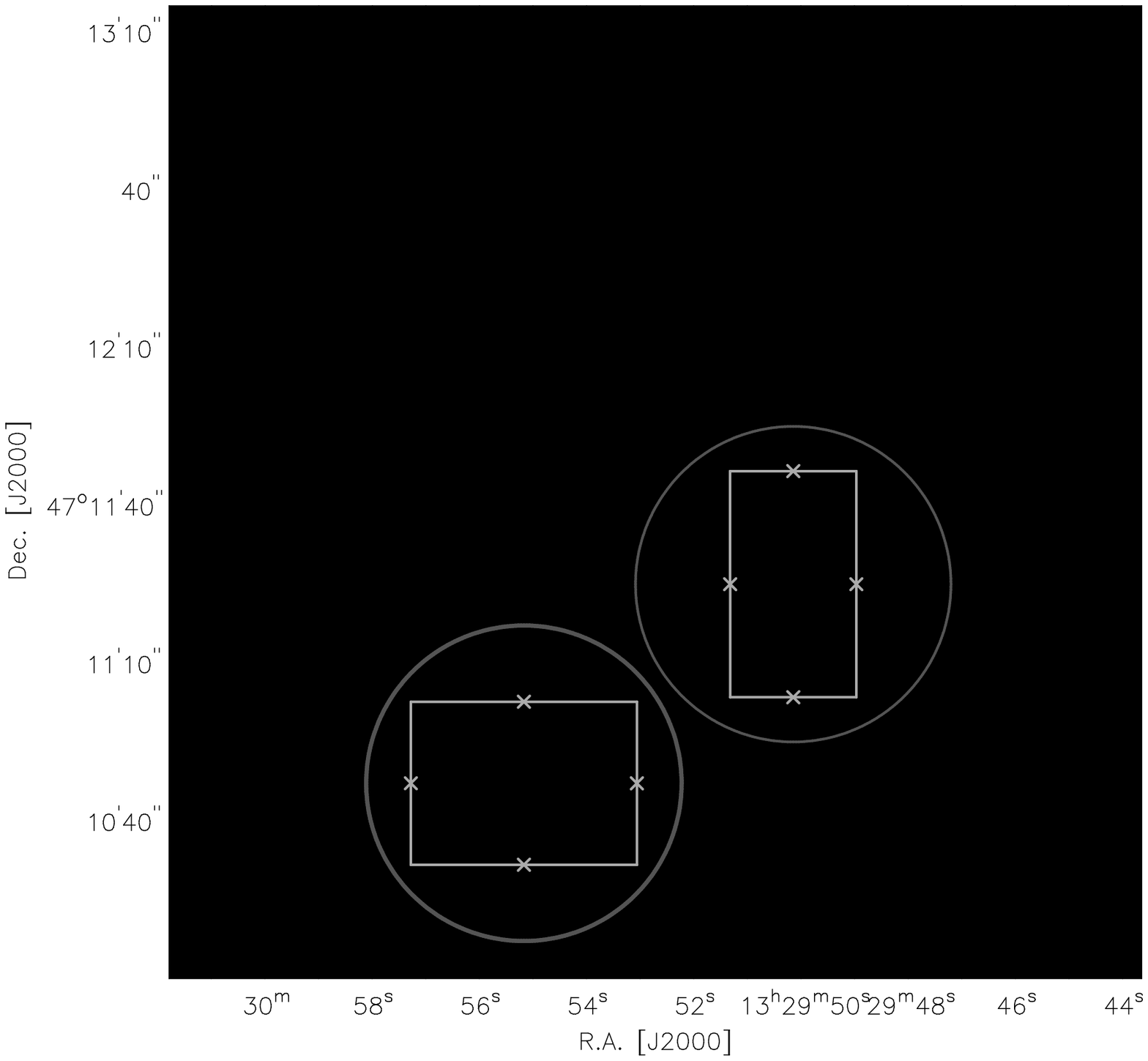}
\caption{Location of the two OVRO pointings (dark circles) overlaid
  onto the integrated $^{12}$CO(1-0) line emission (contours) from
  \cite{aal99} with short spacing correction (SSC) applied. The size
  of the circles corresponds to the primary beam at 3mm. The light
  gray rectangles outline the western (top) and southern region
  (bottom) that are studied.}
\label{fig:point}      
\end{figure}

For this multi-transition study of the molecular gas spiral arms in
M\,51a, two prominent regions (Fig. \ref{fig:point}) were selected
based on the interferometric map of the \coone\ line emission obtained
earlier with the Owens Valley Radio Observatory (OVRO)
mm-interferometer \citep{aal99}: (a) the region containing the
brightest \coone\ peak (hereafter: west), and (b) the region showing
evidence for shear across a spiral arm south of the nucleus
(hereafter: south) comprised of the GMAs A8, A9, and A10
\citep[nomenclature following][]{ran90}. Each region was covered by a
single pointing with the OVRO interferometer, and both regions are in
the M1 spiral arm. The coordinates of the pointing centers are given
in Tab. \ref{tab:poi}. We observed the following key tracers for
molecular gas: \cotwo, \cotre, \coeig, \hcn\ and \hco. In order to
accommodate simultaneous observations of the lines of \cotre\ and
\coeig\ as well as \hcn\ and \hco, the spectral set-up for the southern
and western pointing was centered at slightly different velocities. Tab.
\ref{tab:spec} lists the velocity of the central channel as well as
the resolution adopted for the data analysis. The data were obtained
between January 2001 and March 2002 in the C, L, E, and H
configuration (for details see Tab. \ref{tab:data}). J1159+292 and/or
J1310+323 served as phase calibrators and were observed every 20min
between source observations. For passband calibration 3C273, 3C84,
and/or 3C345 were used. Observations of Neptune or Uranus were
observed in several tracks throughout the observing period and allowed
the flux calibration of the phase calibrator which was when
interpolated for the M51 observations. The uncertainty of the absolute
flux calibration is about 10\% (15\%) at 3mm (1mm). The calibration
of the $uv$ data was done using the OVRO software MMA
\citep{sco94}. For mapping and CLEANing we used the software package
MIRIAD \citep{sau95}. The FWHM of the primary beam of the OVRO
interferometer is about 60$\as$ (38$\as$) at 3mm (1mm).

In addition to the new observations, we use the \coone\ mosaic obtained
by \cite{aal99}. The mosaicked map consists of 19 pointings
observed in the C, L and H configuration with OVRO. The final spectral
resolution is 7.8\,$\kms$ with 33 channels centered at a velocity of
$v_{LSR} = 472\,\kms$. The sensitivity of the naturally (uniform)
weighted data is 19\,mJy\,beam$^{-1}$ (25\,mJy\,beam$^{-1}$) with a
beam of 3.9$\as\,\times$\,3.3$\as$ (2.9$\as\,\times$\,2.1$\as$). A
complete description of the dataset is given by \cite{aal99}.

\subsection{Millimeter Single Dish Observations}
\label{subsec:single}

For a meaningful comparison of the different tracers of the molecular
gas, the interferometric data need to be corrected for missing short
spacings (SSC - short spacing correction). Therefore, we observed the
regions of the western and southern OVRO pointings with the IRAM 30m to
obtain total power data. Around each pointing center a region of
80$\as\,\times\,$80$\as$ was mapped in the On-The-Fly (OTF) mode in
July 2001. The observations were carried out using the AB
receiver combination in two frequency setups covering simultaneously 
the \cotre\ and \cotretwo\ or the  \coeig\ and \cotretwo\ lines 
in dual-polarization mode at 3 and 1mm. Since we did not observe
the \cotretwo\ line at OVRO, the 1\,mm data will not be described
in the following.

The telescope beam size at the 3mm observing frequencies is $22''$.
We used four of the 1\,MHz filter-banks as spectral back-ends. Each
unit has 256 channels resulting in a velocity resolution of 2.6$\kms$
and a total velocity coverage of 670$\kms$ at 3mm, respectively.
System temperatures were typically $\approx$\,170\,K ($T_A^*$) for
both frequency setups. Pointing was checked frequently and was found
to be stable to within $3''$. Calibration was done every 15\,min using
the hot/cold--load absorber measurements. The OTF observations were
done in equatorial coordinates, alternating the two orthogonal
scanning directions. The scan speed was $2''/s$ with a dump time of
$2\,s$ and steps orthogonal to the
scanning direction of $4''$ yielding spectra spaced by $4''$ on sky. Data
were reduced using the CLASS software. Linear baselines were
subtracted from each spectrum and intensities were converted to main
beam brightness temperatures ($T_{\rm mb}$). The calibrated,
baseline-subtracted spectra were used to produce a 3-D data cube.
During the gridding we reduced the velocity resolution to 5$\kms$ for
both lines. The resulting rms noise level for \cotre\ is 9\,mK in
both regions. For the \coeig\ line we achieved an rms noise of 6\,mK
in the west and 12\,mK in the south. We estimate fluxes are accurate
to $\pm$10\% .

For the SSC of the OVRO mosaic of the \coone\ data from \cite{aal99}
and our OVRO \cotwo\ observations, we used the \coone\ and \cotwo\
single dish maps obtained with the NRO 45m telescope
\citep{nak94}\footnote{\tt http://www.nro.nao.ac.jp/~nro45mrt/COatlas/} and the HERA receiver
array at the IRAM 30m telescope \citep{schu07}, respectively. The
\coone\ data cube has spatial resolutions of $16''$ and a spatial
gridding of $7.5''$ on sky for the region of interest here. The
\cotwo\ data cube was observed in OTF (On-The-Fly) mode and has a
spatial resolution of $11''$ and a spatial gridding of 7''. The rms
noise levels at the resampled spectral resolution of 5.3$\kms$ and
4.5$\kms$ is 60\,mK and 30\,mK ($T_{\rm mb}$) for the \coone\ and
\cotwo\ line, respectively. We have checked and corrected the
astrometry of the single-dish maps relative to the OVRO maps by
comparing the peaks in the integrated intensity maps before the short
spacing correction. For all maps the astrometry turned out to be
consistent within the pointing errors of the single-dish observations
with applied shifts of $\rm (\delta RA=4.5'', \delta DEC=-1.5'')$ and
$\rm (\delta RA=3.2'', \delta DEC=0.0'')$ for the \coone\ and \cotwo\ line,
respectively.

\subsection{Short Spacing Correction}

In order to recover the flux resolved out by the interferometer
(missing flux), we combined the OVRO interferometric and single dish
data following the short spacing correction (SSC) method outlined by
\cite{wei01}. In brief, this method uses the cleaned interferometer
map as a starting point and replaces the central part of the {\it
  uv}-plane with visibilities calculated from the single dish map
using the same spatial and velocity grid and flux units as the
interferometer map. The method requires knowledge of the
interferometer clean beam and the single dish beam (which has been
approximated by a Gaussian). The method has no free parameters except
the choice which part of the {\it uv}-plane is replaced by the single
dish data. All visibilities shorter than the smallest projected
baseline of the interferometer data were taken from Fourier transforms
of the single dish data cubes. For the analysis, we produced short
spacing corrected data cubes at 4.5$\as$ (2.9$\as$) for both (west
only) regions. We find that the missing flux in the OVRO pointings is
$\sim 50$, 40 and 20\% for the \coone, \cotre\ and \coeig\ lines,
respectively, and $\sim 70\%$ for the \cotwo\ line. The missing flux
percentage is lower towards the strong emission peaks in the
interferometer maps. All maps presented in the following are the short
spacing corrected versions.

\subsection{HST Archival Data}

For comparison to sites of recent star formation we use archival
HST data from the WFPC2 camera in the V band and H$\alpha$ line as
well as the NICMOS Pa$\alpha$ line as presented by \cite{sco01}.

\section{The molecular ISM}
\label{sec:co}

In order to study the molecular gas emission, we created moment maps
from the short spacing corrected data cubes using the GIPSY (Groningen
Image Processing System) task 'moment'. We used a 3$\sigma$
(5$\sigma$) or 5$\sigma$ (7$\sigma$) limit and required that emission
above this threshold is present in at least two adjacent channels to
minimize the noise for the intensity maps (velocity fields). Note that
this is a very strict cut and low level emission will be missed by the
moment maps presented here.

\subsection{CO Morphology}

\begin{figure}
\centering
\includegraphics[height=17cm,angle=-90]{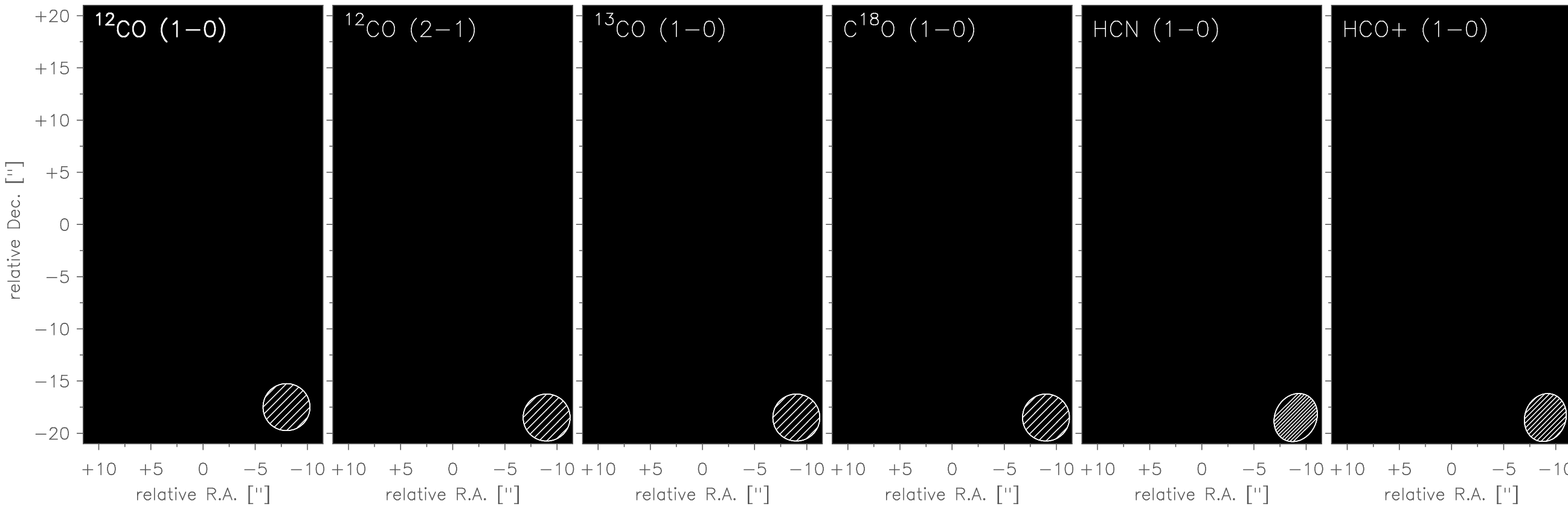}
\caption{The western region at 4.5'' resolution in molecular line
  emission of (from left to right): \coone, \cotwo, \cotre,
  C$^{18}$O\,(1-0), HCN\,(1-0), and HCO+(1-0). All maps shown are made
  from the SSC data cubes. The beam is shown in the bottom right
  corner of each panel. The angular offset is relative to the pointing
  center.}
\label{fig:w_all}      
\end{figure}

\begin{figure}
\centering
\includegraphics[height=18cm,angle=-90]{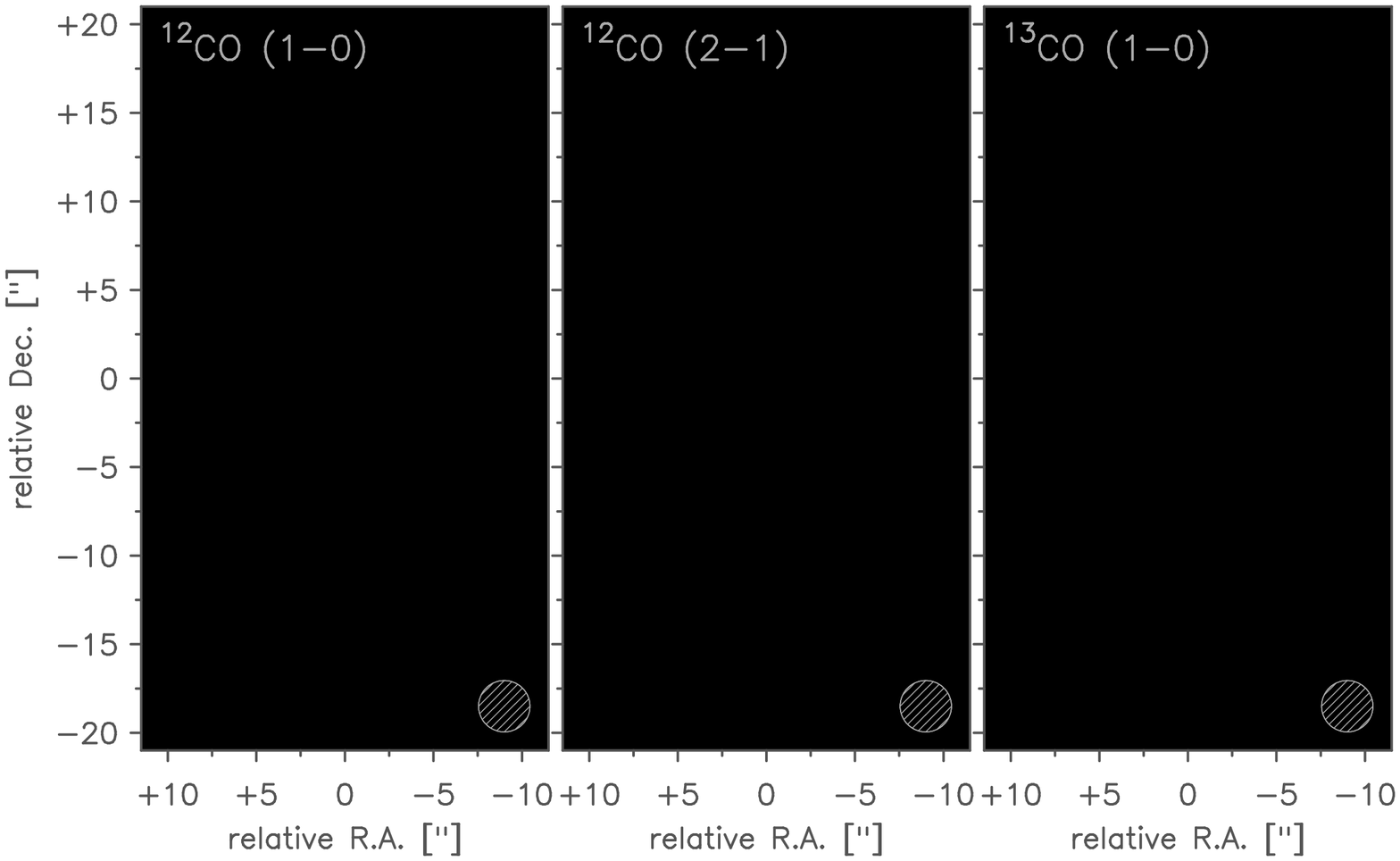}
\caption{The western region at 2.9'' resolution in molecular line
  emission of \coone\ ({\it left}), \cotwo\ ({\it middle}), and
  \cotre\ ({\it right}). All maps shown are made from the SSC data
  cubes. The beam is shown in the bottom right corner of each panel.
  The angular offset is relative to the pointing center.}
\label{fig:w_all2.9}      
\end{figure}

\begin{figure}
\centering
\includegraphics[height=20cm,angle=0]{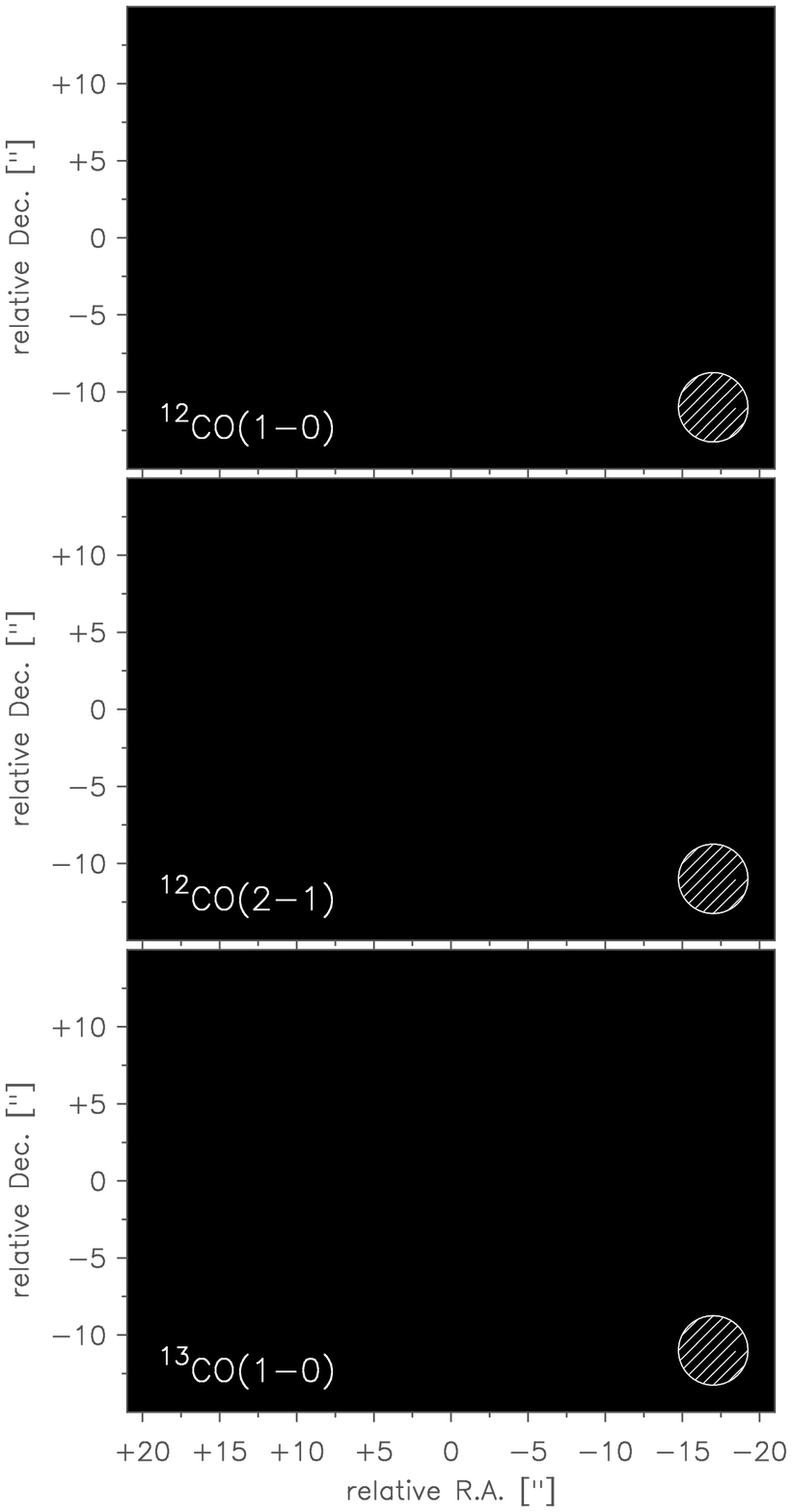}
\caption{The southern region at 4.5'' resolution in molecular line
  emission of \coone\ ({\it top}), \cotwo\ ({\it middle}), and \cotre\
  ({\it bottom}). All maps shown are made from the SSC data cubes. The
  beam is shown in the bottom right corner of each panel.  The angular
  offset is relative to the pointing center.}
\label{fig:s_all}      
\end{figure}

Strong line emission has been detected from the \coone, \cotwo, and
\cotre\ transition while \coeig\ is only detected at a low level in
some smaller peaks (see Fig. \ref{fig:w_all}, \ref{fig:w_all2.9} and
\ref{fig:s_all}). However, the relative brightness of peaks within the
arms is different for each line suggesting that different physical
conditions might be present in different cloud structures. A slight
change in CO morphology is apparent for the western region when
comparing the \coone, \cotwo, and \cotre\ line emission at 4.5'' (Fig.
\ref{fig:w_all}) and 2.9'' (Fig. \ref{fig:w_all2.9}) resolution: the
small-scale geometry like the ridge connecting the peaks at a relative
declination at +5'' and 0'' becomes more evident in the \coone\ and
\cotre\ maps. The difference is likely due to an extended low
excitation component which is contributing more to the \coone\ line
emission than the \cotwo\ one.

\begin{figure}
\centering
\includegraphics[height=18cm,angle=0]{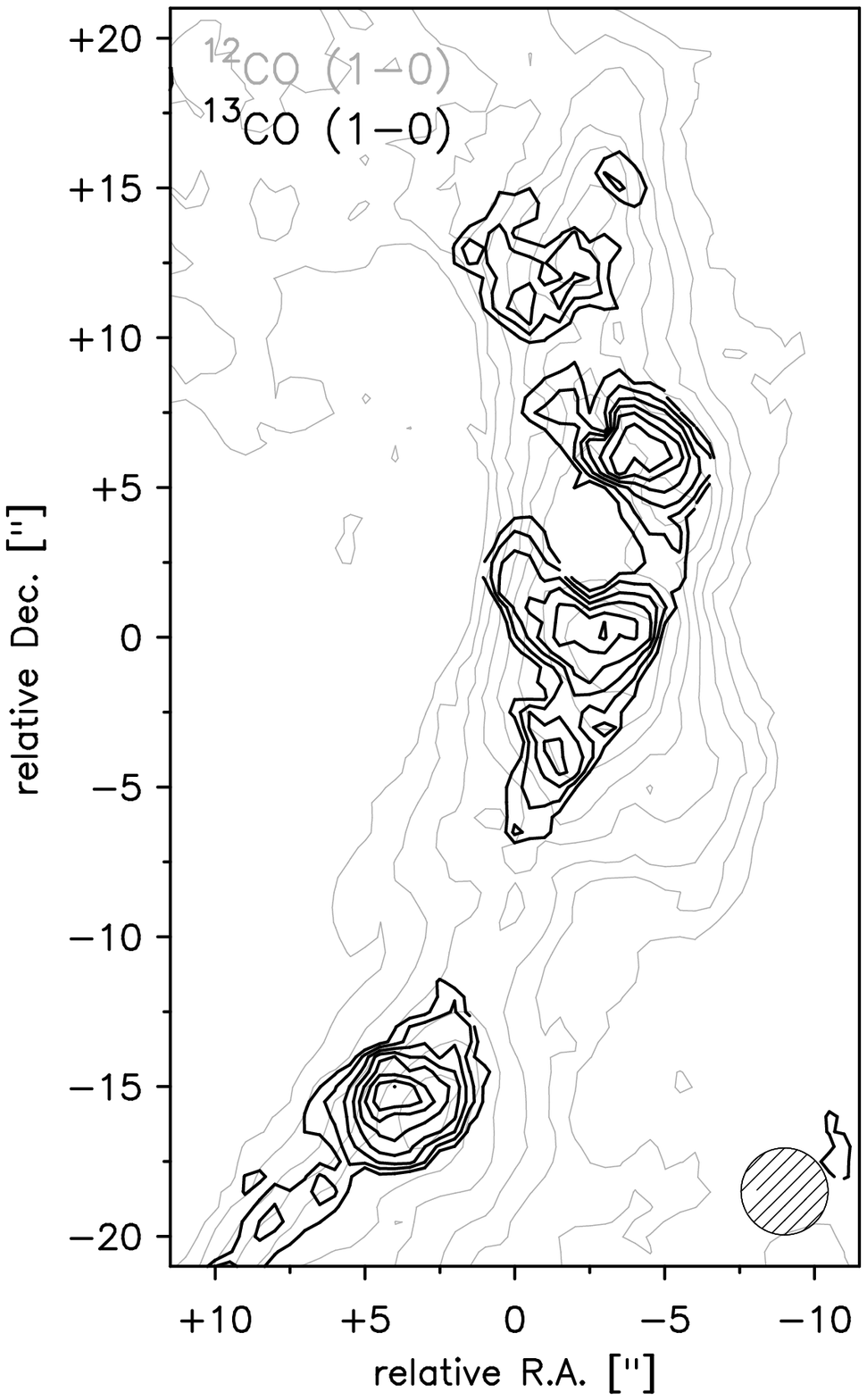}
\caption{Comparison of the location of the peak emission in the
  \coone\ (gray contours) and \cotre\ (black contours) in the western
  arm at 2.9'' resolution. The contours are in steps of 10\% of the
  maximum value of $\rm 295\,K\,km\,s^{-1}$ starting at 20\% for the
  \coone\ map and in steps of 10\% of the maximum value of $\rm
  37.3\,K\,km\,s^{-1}$ starting at 30\% for the \cotre\ data. All maps
  shown are made from the SSC data cubes. }
\label{fig:13co10_w}      
\end{figure}

\begin{figure}
\centering
\includegraphics[height=17cm,angle=-90]{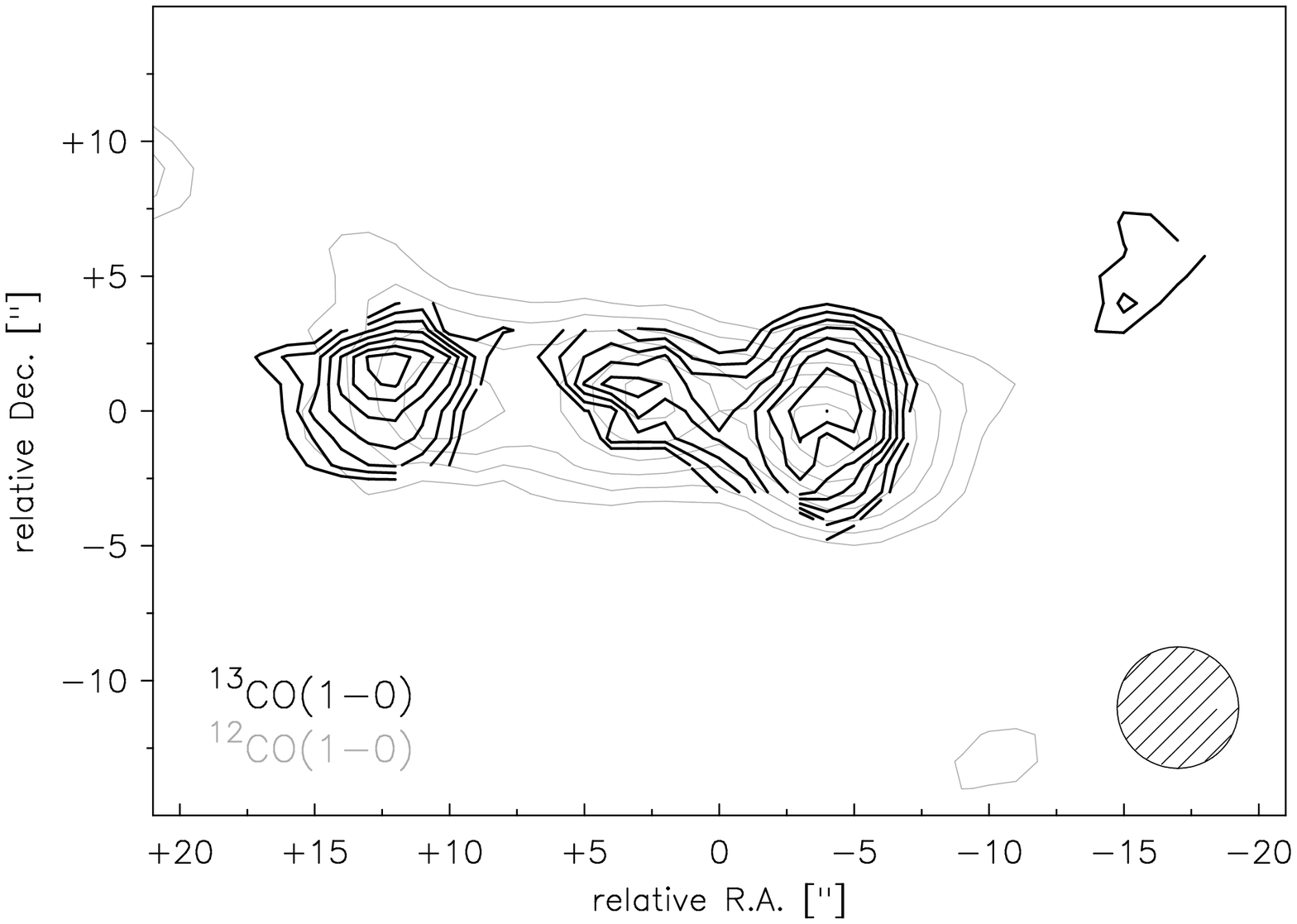}
\caption{Comparison of the location of the peak emission in the
  \coone\ (gray contours) and \cotre\ (black contours) in the southern
  arm at 4.5'' resolution. The contours are in steps of 10\% of the
  maximum value of $\rm 201\,K\,km\,s^{-1}$ starting at 30\% for the
  \coone\ map and in steps of 20\% of the maximum value of $\rm
  5.0\,K\,km\,s^{-1}$ starting at 20\% for the \cotre\ data. All maps
  shown are made from the SSC data cubes. }
\label{fig:13co10_s}      
\end{figure}

No significant offset is seen between the peaks in \coone\ and \cotre\
(Fig. \ref{fig:13co10_w} and \ref{fig:13co10_s}). We verified that the
location of the peaks in the moment 0 maps is not significantly
affected by the inclusion of the single dish data. Small offsets of a
fraction of the beam are seen for some peaks (mainly in the \cotre\
moment maps) as is expected in the case of low S/N data (being boosted
by the additional signal from the single dish data). In addition, as
the physical conditions across the spiral arms are changing as shown
in \S \ref{sec:lvg}, a perfect one-to-one correspondence between the
\coone\ and \cotre\ morphology is not expected. The absence of an
obvious shift is in apparent contradiction to the result of
\cite{tos02} who found that the \cotre\ emission is leading the
\coone\ emission in particular in the southern region. One explanation
could be that this apparent offset is caused more by an offset in the
diffuse gas than in the more clumpy components which are better
resolved with the OVRO data. Another possibility is the instrumental
offset we had to correct for when aligning the single dish and
interferometric data (see \S \ref{subsec:single}). As the offset is
mainly in RA for the \coone\ single dish data this could explain the
observed difference seen for the Western arm. If an instrumental effect
is also causing the mismatch for the Southern arm, the astrometry of
the \cotre\ single dish data should be responsible.

The smooth ridges present in the \coone\ intensity map of \cite{aal99}
break up into smaller peaks at $\sim$1.6'' resolution achieved in the
\cotwo\ data. In particular the western region (Fig. \ref{fig:mom_w})
splits into two gas lanes over a length of about 10'' ($\sim$ 410\,pc)
while the southern region (Fig.  \ref{fig:mom_s}) shows up to three
spurs along the leading side of the spiral arm similar to the features
studied by \cite{cor08}. The peaks in both regions appear to be
slightly resolved at that resolution.

\subsection{CO Kinematics}

\begin{figure}
\centering
\includegraphics[height=17cm,angle=-90]{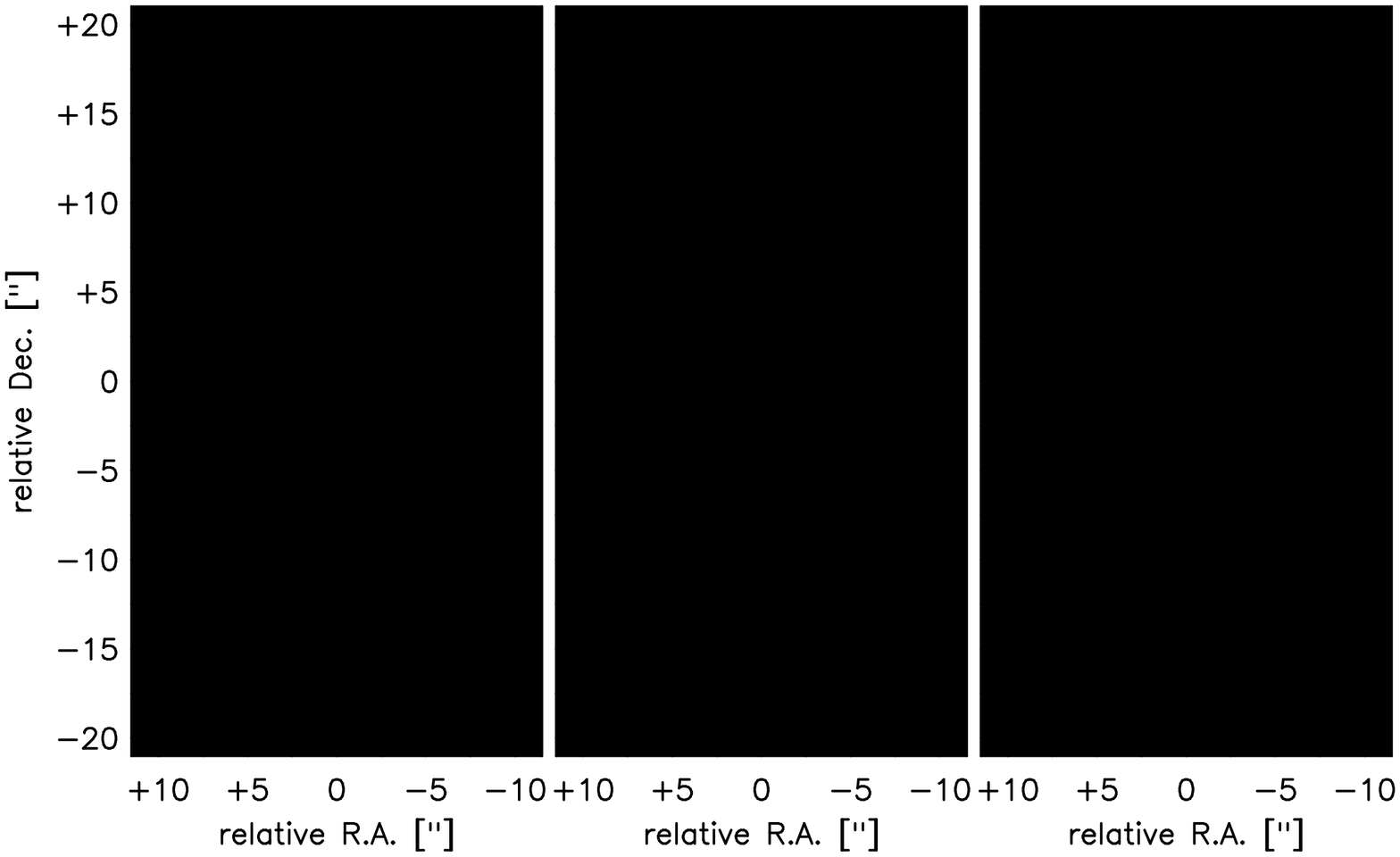}
\caption{Moment maps of the \cotwo\ line emission
  (at a resolution of 1.7''$\times$1.5'' and short spacing corrected):
  integrated intensity map ({\it left}), velocity field ({\it
    middle}), and dispersion map ({\it right}). The contours in the
  intensity map are in steps of 10\% of the peak value of $\rm
  23.3\,Jy\,beam^{-1}km\,s^{-1}$, the iso-velocity contours start at
  450\,km/s with steps of 5\,km/s, while the contours in the
  dispersion map start at 5\,km/s with a step of 5\,km/s. The beam is
  shown in the bottom right corner of each panel.  The angular offset
  is relative to the pointing center.}
\label{fig:mom_w}      
\end{figure}

\begin{figure}
\centering
\includegraphics[height=18cm,angle=0]{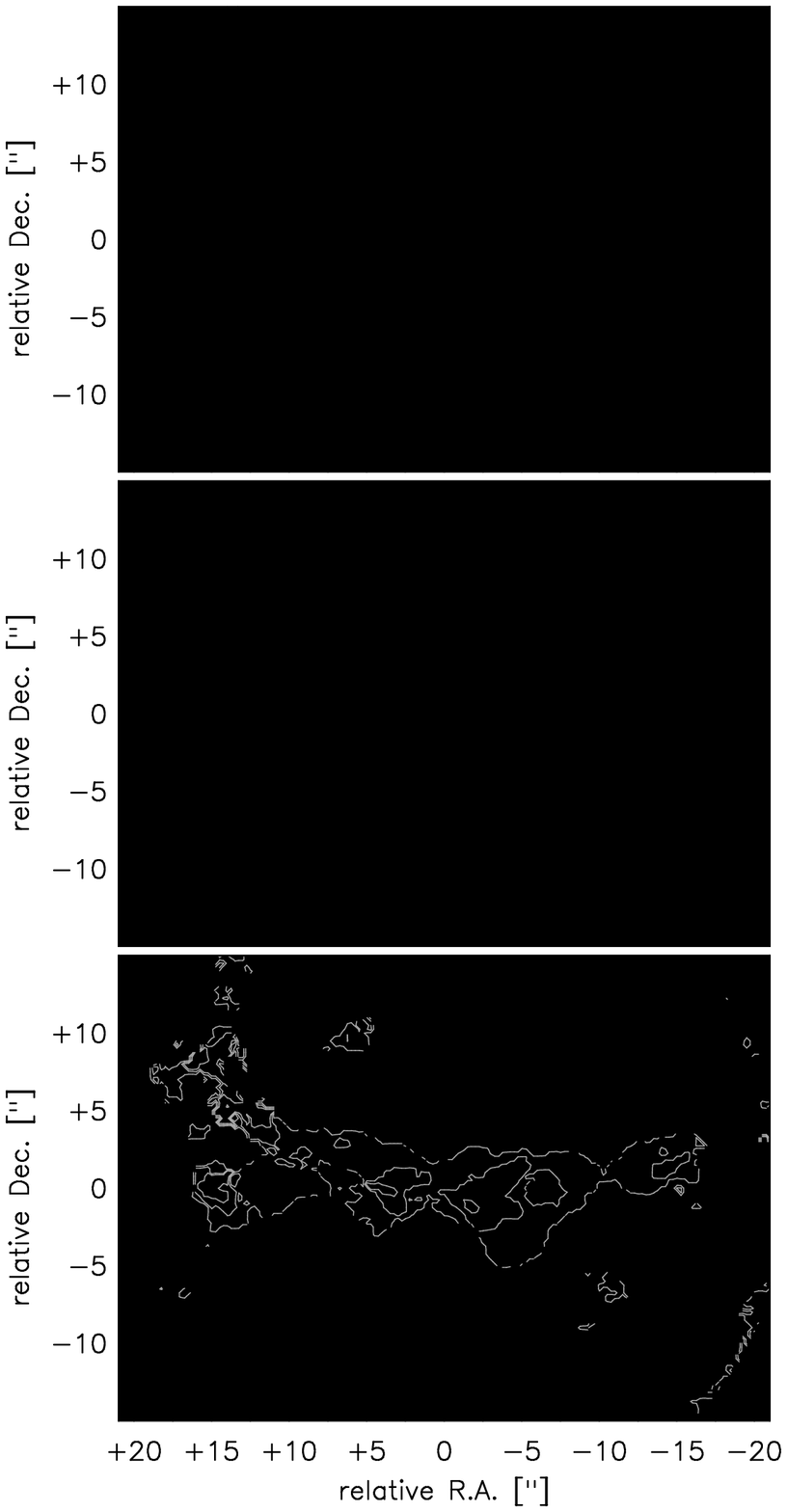}
\caption{Moment maps of the \cotwo\ line emission (at a
  natural resolution of 1.7''$\times$1.5'' and short spacing
  corrected): intensity map ({\it top}), velocity field ({\it middle}),
  and dispersion map ({\it bottom}). The contours in the intensity map
  are in steps of 10\% of the peak value of $\rm
  12.4\,Jy\,beam^{-1}km\,s^{-1}$ starting at 30\%, the iso-velocity
  contours start at 490\,km/s with steps of 10\,km/s, while the
  contours in the dispersion map start at 5\,km/s with a step of
  5\,km/s. The beam is shown in the bottom right corner of each panel.
  The angular offset is relative to the pointing center.}
\label{fig:mom_s}      
\end{figure}

The 1.6'' resolution \cotwo\ data allows us to study the molecular gas
kinematics at about 65\,pc resolution. Although the overall kinematics
corresponds well with the low angular resolution velocities of
\cite[][their Fig. 3]{aal99}, the iso-velocities showing small
deviations indicate that some GMC associations (GMAs) move faster or
slower (fig. \ref{fig:mom_w} middle). While the general deviations
from a pure rotational disk are present as well indicating an overall
non-circular component, some peaks stand out in the velocity
dispersion map suggesting that a few peculiar velocities in the
velocity maps are due to a superposition of GMCs rather than actual
streaming motions. The most obvious example is the peculiar velocity
field structure identified as streaming motion across the southern arm
by \cite{aal99} that is likely caused by a superposition of two GMCs
rather than a distinct velocity feature due streaming (Fig.
\ref{fig:mom_s} middle and bottom). The high velocity dispersion
observed in the GMA at offsets of +5'' and +0'' \citep[GMA A10
following the nomenclature of ][]{aal99} and the break-up into several
sub-components of the GMA at offsets +12'' and +2'' (GMA A9) support
our interpretation. In addition, the 1.6'' resolution \cotwo\ data
reveals that the higher velocity dispersion ridge seen in the \coone\
mosaic of \cite{aal99} in particular in the western arm belongs to the
second set of line peaks on the inner side of the spiral arm (see Fig.
\ref{fig:mom_w} right).

\subsection{Dense Gas Tracers}

HCN(1-0) and HCO+(1-0) line emission was only detected in the western
arm region at a low significance level (see Fig.  \ref{fig:w_all}).
HCO+(1-0) is confined to the bright CO peak at about -2'' and +0''
while HCN(1-0) might be associated also with another bright CO peak
north of this position. Given the low significance of the detection
these data are not included in the following analysis.

\section{Non-LTE Analysis}
\label{sec:lvg}

In order to estimate the physical conditions in the molecular gas
along the gas spiral arm we measured the CO line emission in all four
transitions for several positions along the spiral arms. As emission
from the \coeig\ line was only detected in a small fraction of the
positions, we will focus on the \coone, \cotwo, and \cotre\ line
measurements for our analysis.

\subsection{Line Ratios}

\begin{figure}
\centering
\includegraphics[height=18cm,angle=-90]{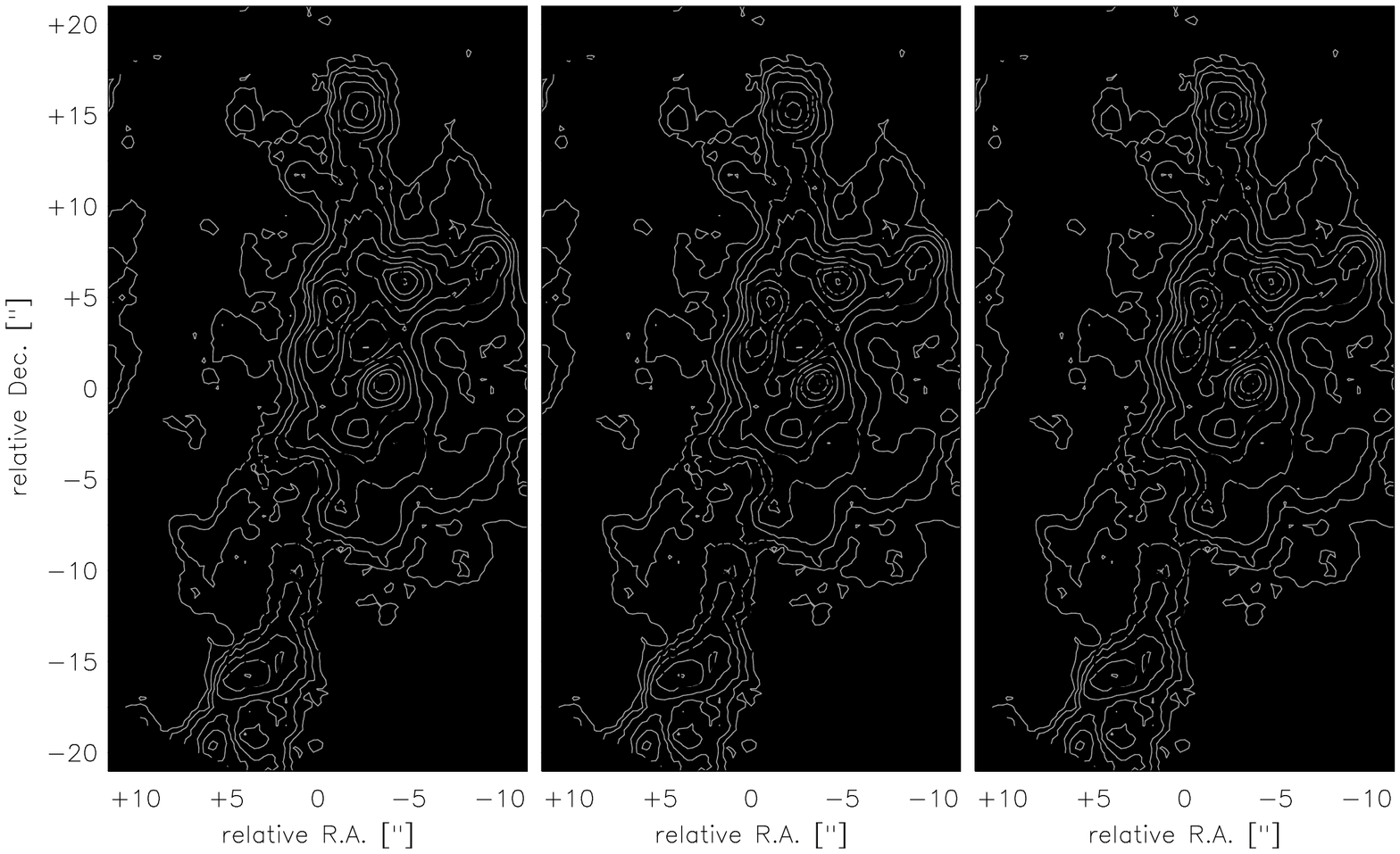}
\caption{Location of the positions studied in the western arm.
  Numbering of the positions is given in the {\it left} panel where
  the size of the circles corresponds to a resolution of 2.9''. The
  derived line ratios R$_{21}$ of the \cotwo\ and \coone\ emission and
  R$_{13}$ of the \cotre\ and \coone\ emission for the individual
  positions is listed in the {\it middle} panel. Positions with empty
  circles have only upper limits in at least one line. The derived
  kinetic temperature $\rm T_{kin}$ and volume density of $\rm H_2$
  n($H_2$) from the LVG analysis are given in the {\it right} panel.
  The integrated \cotwo\ intensity (contours) is overlaid on the
  \coone\ intensity (gray-scale).}
\label{fig:rat_w}      
\end{figure}

\begin{figure}
\centering
\includegraphics[height=18cm,angle=0]{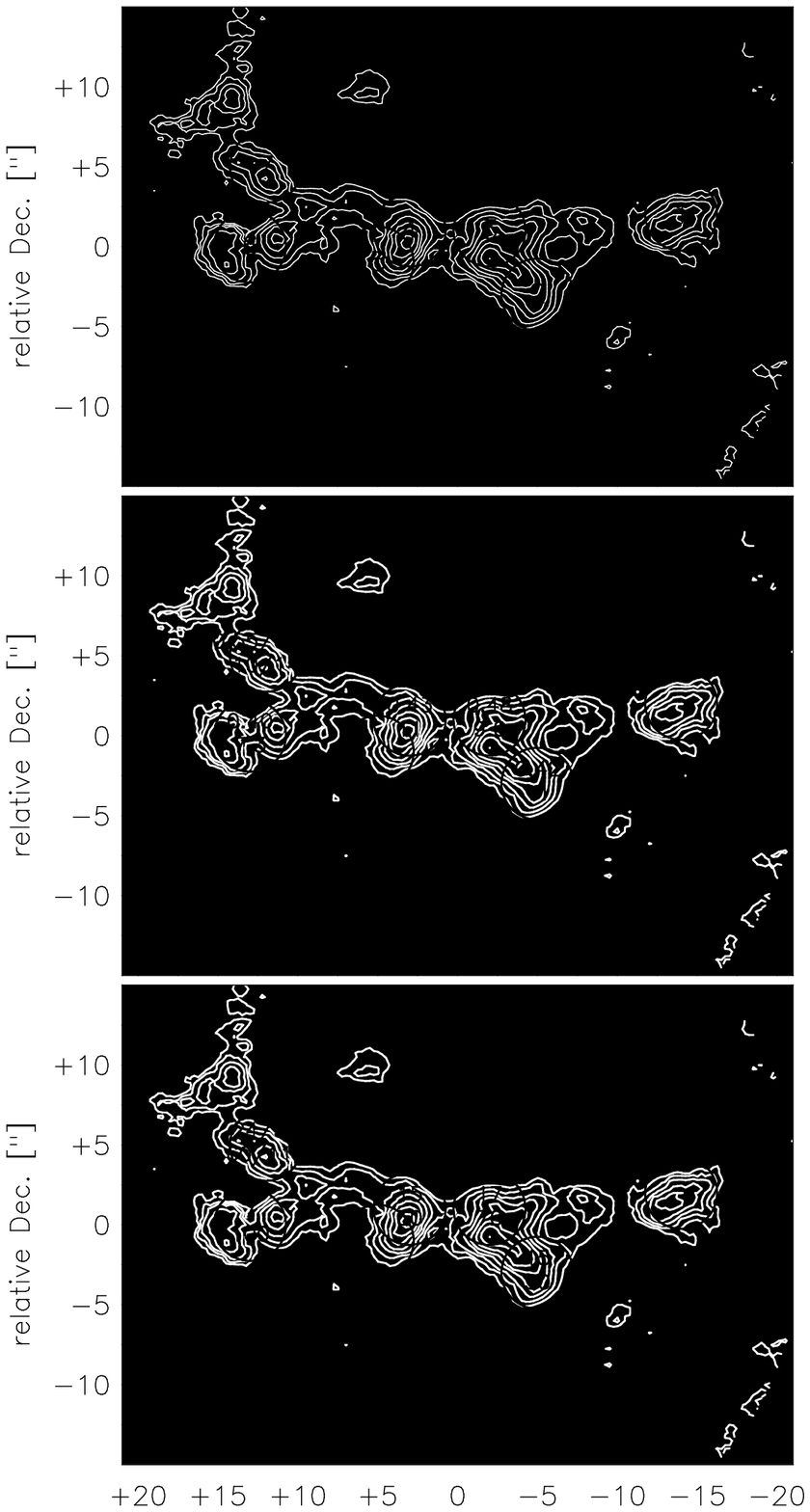}
\caption{Location of the positions studied in the western arm.
  Numbering of the positions is given in the {\it top} panel where the
  size of the circles corresponds to a resolution of 4.5''. The
  derived line ratios R$_{21}$ of the \cotwo\ and \coone\ emission and
  R$_{13}$ of the \cotre\ and \coone\ emission for the individual
  positions is listed in the {\it middle} panel. Positions with empty
  circles have only upper limits in at least one line. The derived
  kinetic temperature $\rm T_{kin}$ and volume density of $\rm H_2$
  n($H_2$) from the LVG analysis are given in the {\it bottom} panel.
  The integrated \cotwo\ intensity (contours) is overlaid on the
  \coone\ intensity (gray-scale).}
\label{fig:rat_s}      
\end{figure}

14 and 9 positions were selected in the western arm region by
basically centering them on distinct peaks of the CO line emission in
the 2.9'' and 4.5'' resolution maps, respectively. In the southern
region 7 positions were selected in a similar manner at 4.5''
resolution only. The western 2.9'' resolution and southern 4.5''
resolution positions are indicated in Fig. \ref{fig:rat_w} and
\ref{fig:rat_s}, respectively. For the measurement of the line ratios
$R_{21}$ of the \cotwo\ to \coone\ line, $R_{13}$ of the \coone\ and
\cotre\ line and $R_{18}$ of the \coone\ and \coeig\ line all data
cubes were converted from flux density with units of $\rm Jy/beam$ to
brightness temperature with units of $\rm K$. For each position
spectra of the four CO lines were extracted at the central pixel.  In
a few positions spectra showed evidence for two line
components suggesting that more than one GMC is contributing to the
emission peak. In these cases two Gaussians were fit to the spectra.

The integrated line fluxes were derived from Gaussian fits to the
spectra and are listed in Tab. \ref{tab:flux} and Tab. \ref{tab:fluxa} for
the 2.9'' and 4.5'' resolution data. Upper limits for the line flux
were derived using three times the rms in the spectrum multiplied by
the FWHM measured for the \coone\ line.  The listed line width
corresponds to the FWHM of the \coone\ line. The line ratios $R_{21}$,
$R_{13}$ and $R_{18}$ were derived by scaling the overall shape of the
\cotwo, \cotre\ and \coeig\ spectra to the corresponding \coone\
spectrum and are listed in Tab. \ref{tab:flux} and \ref{tab:fluxa} as
well. Thus the listed line ratios might not agree perfectly with a
line ratio derived from the integrated line fluxes. This approach is
more robust in the case of low S/N spectra than using the peak or
integrated line flux which could be strongly effected by noise peaks.
If no line was detectable in the data, we used the peak flux (upper
limit) of the lines in question to derive a lower limit for the
$R_{13}$ and $R_{18}$ ratios.

The derived line ratios $R_{21}$ and $R_{13}$ for the 2.9'' resolution
western and 4.5'' resolution southern region are shown as a function
of position in Fig. \ref{fig:rat_w} and \ref{fig:rat_s}, respectively.
In the western region the $R_{21}$ ratio is basically monotonically
decreasing as a function of galacto-centric distance with a value of
$R_{21}$=0.72 at position w1 corresponding to a radius of about
1.1\,kpc ($\sim$ 27'') and values between 0.42 and 0.47 for positions
w11 to w14 which are at a radial distance of about 1.5\,kpc ($\sim$
37''). The $R_{13}$ ratio on the other hand shows a spread of values
between 4.5 (w14) and 7.5 (w5) with no clear radial trend.  However,
most of the higher values are found close to the area of brightest CO
emission. While the western region covers a larger range in galactic
radii, the southern region is basically tracing molecular gas located
at roughly 2.6\,kpc distance from the center. Thus the $R_{21}$ and
$R_{13}$ have fairly similar values over most of the positions, except
for the most eastern ones (S6 and S7) which are located close to a
very large HII region (see \S \ref{sec:sf}) and show a double line
profile. It is interesting to note that both regions exhibit one
component where the ratios are slightly higher than those found for
the southern most positions in the western region. We note that all
trends seen within a single pointing should be real, as calibration
uncertainties should only affect the absolute flux levels and thus all
positions investigated in the same way.

\subsection{LVG Parameters}

To constrain the physical conditions of the GMCs we use an spherical,
isothermal one-component large velocity gradient (LVG) model
\citep{sco74,gol74}. This treatment of the radiative transfer is valid
for clouds in which a large-scale velocity gradient effectively makes
the radiative transfer a local problem not requiring self-consistent
solutions for the entire cloud. The assumption for this model is that
due to the velocity gradient present the emission of a molecule is
Doppler-shifted with respect to molecules in the rest of the cloud.
For the models presented here we fixed the $\rm ^{12}CO$ abundance
relative to H$_2$ and the $\rm ^{12}CO/^{13}CO$ abundance ratio to
8.0e-5 and 30, respectively. The numbers correspond to values found in
GMCs in the inner 3\,kpc of the disk of the Milky Way
\citep[e.g.][]{lan93,mil05}. LVG models were calculated for two
different assumption on the velocity gradient: a) a fixed value of
$\frac{\rm d\it{v}}{\rm d\it{r}}$ set to $\rm 1\,km\,s^{-1}pc^{-1}$
typically for local clouds and reasonable for M51's
clouds\footnote{For an average line width of $\rm 40\,km\,s^{-1}$ (see e.g.
  Tab. \ref{tab:flux}) and an average cloud complex size of
  $\sim$50\,pc (see \S \ref{sec:sf}).}, b) assuming a virialized cloud
such that $\frac{\rm d\it{v}}{\rm
  d\it{r}}=3.1\sqrt{\frac{n(H_2)}{10^4}}$ \citep[e.g.][]{gol01}. LVG
line intensity ratios were computed for kinetic gas temperatures $\rm
T_{kin}$ between 3 and 200\,K and H$_2$ densities log\,$n(\rm H_2)$
between 2.0 and 4.0 \citep[for details of the model, see][]{wei01}.

\subsection{M51 model results}

The results of the LVG analysis are presented in Fig. \ref{fig:rat_w}
and \ref{fig:rat_s} and summarized in Tab. \ref{tab:lvg}. Typical
relative errors on the derived temperature are of the order of
(75-100)\% implying that solutions preferring higher temperatures are
in principle also consistent with the low temperatures found for
several locations. For the H$_2$ densities of $N(\rm H_2)$ and $n(\rm
H_2)$ we find typical relative errors of about 25\% (going up to
$\sim$50\% in the less constrained cases). These errors are
independent on the assumptions made for the velocity gradient.

The bulk of the CO emission probed here arises from cold clouds
(T$_{kin} \sim$ 16-20\,K, depending on the assumption made for the
velocity gradient) with moderate H$_2$ density of $n(\rm H_2) \approx
240-120\,cm^{-3}$.  It is interesting to note that the H$_2$ density
in the assumption of a virialized cloud is typically (50-65)\% of the
H$_2$ density derived for a fixed velocity gradient. The average
conversion factor derived from the LVG analysis of $X_{CO} = \rm
(1.3-2.0)\times 10^{20}\,cm^{-2}K^{-1}km^{-1}s$ is close to the
Galactic value\footnote{First determined by \cite{san84}.} of $X_{CO}
= \rm (1.8\pm0.3)\times 10^{20}\,cm^{-2}K^{-1}km^{-1}s$. This value
was recently derived for the Milky Way by \cite{dam01} suggesting that
the GMCs analyzed here are similar to those in the Milky Way,
especially when assuming that they are all virialized clouds. Thus the
data presented here mainly probe the diffuse component of the ISM and
not the dense gas that is involved in current star formation. In the
remainder we refer to the LVG results assuming virialized clouds.
However, all trends discussed are present in both scenarios.

The most obvious spatial variation of the gas physical condition is a
decrease of the kinetic temperature with increasing galacto-centric
radius along the western spiral arm region. Temperatures decrease from
about 27\,K at 1.1\,kpc (position w1) to $<$10\,K at 1.6\,kpc
(position w14). This might indicate an effect of the interstellar
radiation field (ISRF) on the cloud properties as these radii probe
the transition from the central (bulge) region into the disk (for a
discussion of the effect of HII regions see \S \ref{sec:sf}). In the
southern region, with an almost constant galacto-centric radius of
$\sim$ 2.6\,kpc, no such gradient is detected. These more distant
regions are on average 5-10\,K warmer than the furthest regions in the
Western arm (position w12 and w14). The derived radial temperature
profile by \cite{meij05} used to model the heating mechanism at
850$\mu$m shows a continuous decline of temperature with radius,
however, the change over the distances relevant here is less than
2\,K. The dust temperature map of \cite{ben08} derived by fitting a
grey-body thermal emission spectrum to data at 70, 160, 350 and
850\,$\mu$m shows also an enhanced dust temperature at the center with
a decline for larger radii. However, the dust temperature in the
spiral arms themselves is higher than in the interarm regions. Overall
the change in temperature is not dramatic. \cite{ben08} report a mean
temperature of (25$\pm$3)\,K for the disk while the inner 3\,kpc
region has a higher temperature of 31\,K. Therefore it appears that
more local changes (i.e. within the spiral arm) of the ISRF than the global
radial trend could be causing the apparent observed trend in the
Western arm. However, higher resolution data such as observations from
the Herschel Space Observatory and subsequent detailed modeling are
required to address this in more detail.

There is no obvious large difference in the temperature and H$_2$
density of GMCs close to or around star forming regions as traced by
the presence of H$\alpha$ emission (see \S \ref{sec:sf}, also
indicated in Tab. \ref{tab:lvg}). This finding suggests that the CO
emission arising from or close to star-forming regions is still
dominated by cold gas that is surrounding the HII regions and not
immediately affected by the enhanced radiation from the HII region.
There might be a slight effect present as half of the positions in the
Southern arm encompass HII regions (traced by H$\alpha$ emission) and
show slightly higher temperatures. The impact of star formation on the
molecular clouds (e.g. heating at the cloud surfaces) is not
measurable at a linear resolution of 180\,pc (or even for the higher
120\,pc resolution data). Only two positions stand clearly out: w5 in
the western region and S7(B) in the southern region that show the
highest kinetic temperature of $\rm T_{kin}=50$\,K, average H$_2$
densities and the lowest conversion factor of $X_{CO} = \rm 0.5(0.8)\times
10^{20}\,cm^{-2}K^{-1}km^{-1}s$. Both GMC complexes are located on the
inner edge of the CO spiral arm suggesting that they either live in a
different environment than the clouds studied in the other locations
or that their ISM has been heavily affected by the ongoing neighboring
massive star formation [in the case of S7(B)].

\subsection{$X_{CO}$ measurements for M51}

Several studies were conducted in the past to derive the conversion
factor $X_{CO}$ for the molecular gas in M51. Most of these studies
did find values well below a Galactic value of $X_{CO} = \rm
(1.8\pm0.3)\times 10^{20}\,cm^{-2}K^{-1}km^{-1}s$ \citep{dam01}.
Several reasons could explain the apparent mismatch between our
results and these earlier works. One main advantage of the work
presented here is its superior spatial resolution compared to the
previous results.

Using single dish observations of $^{12}$CO and $^{13}$CO line
transitions and three different methods \cite{gar93} find a value of
$X_{CO} = \rm (0.8\pm0.3)\times 10^{20}\,cm^{-2}K^{-1}km^{-1}s$ for
the center and the molecular spiral arms from LVG modeling and an even
lower value when using only the $^{13}$CO emission line data. Assuming a
constant ratio between the extinction and the $H_2$ column density
$N(H_2)$ as found for our Galaxy \citep{boh78} gave a Galactic value
of $X_{CO} \sim \rm 1.8\times 10^{20}\,cm^{-2}K^{-1}km^{-1}s$.
Extending this work \cite{gue95} added 1mm continuum measurements to
estimate $X_{CO}$ via the gas-to-dust ratio, again getting factors
about 4 times lower than the standard value. The most likely
explanation for the mismatch is that in these studies individual GMC
complexes are not resolved and thus a significant mixing of vastly
different cloud properties can happen.

Using interferometric observations of the molecular gas disk \cite{nak95}
applied the $A_V$ method\footnote{The adopted method by \cite{nak95}
  takes into account gas located behind HII regions on a statistically
  basis.} to those HII regions in the spiral arms that had both
H$\alpha$ as well as radio continuum emission \citep{vdhul88} that
allowed for a good extinction estimate. They obtained a value of
$X_{CO} \sim \rm (0.9\pm0.1)\times 10^{20}\,cm^{-2}K^{-1}km^{-1}s$ at
about 10'' resolution. Among possible sources of error they list a
bias for probing gas with higher excitation temperatures as they are
looking toward HII region where the gas could be potentially more
heated. They also caution that the conversion factor in the inter-arms
might be completely different as the conditions for the molecular gas
are expected to vary as well. Given that most of the gas for which we
derived the conversion factor is not close to HII regions (see \S
\ref{sec:sf}) those measurements are very likely affected by probing a
very different environment than we did.

Interestingly, applying virial mass measurements to the GMAs in the
spiral arms of M51, \cite{ran90} and \cite{adl92} come to very
opposite conclusions of $X_{CO} \sim \rm 3\times
10^{20}\,cm^{-2}K^{-1}km^{-1}s$ and $X_{CO} \sim \rm 1.2\times
10^{20}\,cm^{-2}K^{-1}km^{-1}s$, respectively. Both studies measure
virial masses assuming that the GMAs are resolved with typical sizes
of 300-400\,pc. In order to apply this method the assumption must be
that the GMAs are bound complexes which might not be true given the
fact that it is unclear whether they are just a gas accumulation due
to orbit crowding or indeed self-gravitating entities
\citep[e.g.][]{ran90,adl92}. This is still an open issue discussed by
\cite{kod09} for the GMAs identified in their high fidelity map as the
elongated structure could easily be interpreted as being due to
orbit crowding rather than resembling a bound entity even at their 4''
resolution. As our high resolution \cotwo\ data shows that these GMAs
break up into several clumps that might no longer even correspond to a
single gas lanes (as e.g. seen in the Western region), it seems
sensible that the assumption of bound systems is not valid for all
GMAs. In addition, enhanced velocity dispersion due to shocks or
streaming motions will critically affect the measurements of the line
width at the 10'' resolution of the data used by the previous studies.
Since \cite{adl92} mainly analyzed GMAs located in the central region
where the dynamical properties are changing faster, this could be a
possible explanation for the different results of both works as the
method used is basically the same.

Taking a different approach and combining measurements of the first three
transitions of $^{12}$CO with those of ISM cooling lines, \cite{kra05}
used a PDR (photon-dominated region) model. However, their derived
densities for the center and two regions covering the spiral arms at
slightly larger radii are about 1 order of magnitude higher while
their temperatures of 12.5 and 15 K are similar to our findings. A
recent similar study of the central 3.5\,kpc using a refined PDR model
by \cite{bel07} finds a $X_{CO}$ of 1/10 the Galactic value. As both
studies were also done at low resolution, it is not clear that the ISM
cooling lines as well as the CO lines have the same distribution and
filling factors which could affect the estimation of $X_{CO}$. Given
that the molecular gas studied here is clearly not located closely to
HII regions, it is it not obvious that the PDR assumption is
applicable for our regions (see \S \ref{sec:sf}).

Since it is expected that $X_{CO}$ is changing as a function of
environment, the assumption of a single excitation and a simple
geometry over a large region can frequently yield misleading results.
As our measurements have the highest spatial resolution and are
confined to mainly probing a single environment, they should present
a good estimate of $X_{CO}$ for the spiral arms probed. Given the large
uncertainties involved when averaging over large areas in a galaxies,
they are consistent with the previous results and only highlight the
fact that deriving an average $X_{CO}$ for an entire galaxy is not
straight-forward. In particular the relation between the conversion
factor and the gas density and temperature $X_{CO} \sim \rm
T_{kin}^{-1}n(H_2)^{0.5}$ can explain why the conversion factor is
lower for studies that include a large fraction of the (presumably)
less dense inter-arm gas.

\subsection{Comparison to GMCs in other spiral galaxies}

The LVG models suggest that the physical conditions in the GMCs in the
spiral arms of M\,51 are very similar to those in the Milky Way which
have typical gas temperatures and densities of 10\,K and above
200\,cm$^{-3}$, respectively \citep[e.g.][]{sco87} \citep[for a recent
compilation see][]{tie05}. Also the observed average H$_2$ column
density in M51's GMCs of N(H$_2)=(2.3-3.2)\times10^{22}$\,cm$^{-2}$
is similar to that typically found for GMCs in the Galactic plane of
N(H$_2$) $\sim$ 4$\times$10$^{22}$\,cm$^{-2}$ \citep{sol79}.

Studies of GMCs in local group galaxies found that the cloud
properties do vary between clouds with and without the presence of
star formation. \cite{wil97} analyzed seven GMCs in M33 at 22'' resolution
(corresponding to roughly 90\,pc) using multi-line CO observations in
conjunction with a LVG model. They find that GMCs without star
formation are typically colder with $\rm T_{kin} \sim 10-20\,K$ than
GMCs with star formation present showing $\rm T_{kin} \sim 30 -
100\,K$ for our assumed $\rm [^{12}CO]/[^{13}CO]$ abundance ratio of
30. While the temperatures for the cold GMC population is similar to
our values for M51, their derived densities are about 1 order of
magnitude higher than our estimates for M51. Interestingly, they note
that the sphere of influence for HII regions seems limited as GMCs within
120\,pc distance of an HII region do already show normal properties
and no effect of heating. A recent study of GMCs in the Large
Magellanic Cloud by \cite{min08} also find a correlation between the
star formation activity (measured by the H$\alpha$ flux) and the
temperature and density of the GMCs themselves. They propose an
evolutionary sequence from cool ($\rm T_{kin} \sim 10 - 30\,K$) and
low density (log$\rm (n(H_2)) < 3$) GMCs, through warm ($\rm T_{kin}$
$\sim$ 30 - 200\,K) and cool low density GMCs to warm and dense GMCs
where the first two types are in a young star formation phase where
density has not yet reached high enough values to cause active massive
star formation. GMCs in the last category are in a later star
formation phase where the average density is higher.

Thus we conclude that the bulk of the GMCs observed in M51's spiral
arms are very similar to the ones in our own Galaxy and given their
low temperatures are likely not impacted by (ongoing) star formation.
This is not that surprising as most HII regions are located close to
CO peaks but rarely do coincide (see \S \ref{sec:sf}) and as suggested
by the M33 results from \cite{wil97} the sphere of influence of an HII
region is small. This could also suggest that GMCs located in spiral
arms are different from GMCs located in the center of starburst
galaxies such as M82. \cite{wei01} found significantly higher kinetic
temperatures of $\rm T_{kin}$ between 50 and 190\,K as well as gas
densities $\rm (n(H_2)$ between 5$\times$10$^2$ and
2$\times$10$^4\,\rm cm^{-3}$) and concluded that the ongoing massive
star formation is significantly altering the cloud properties.

\section{The Relation of the Molecular ISM to Sites of Star Formation}
\label{sec:sf}

\begin{figure}
\centering
\includegraphics[height=18cm,angle=-90]{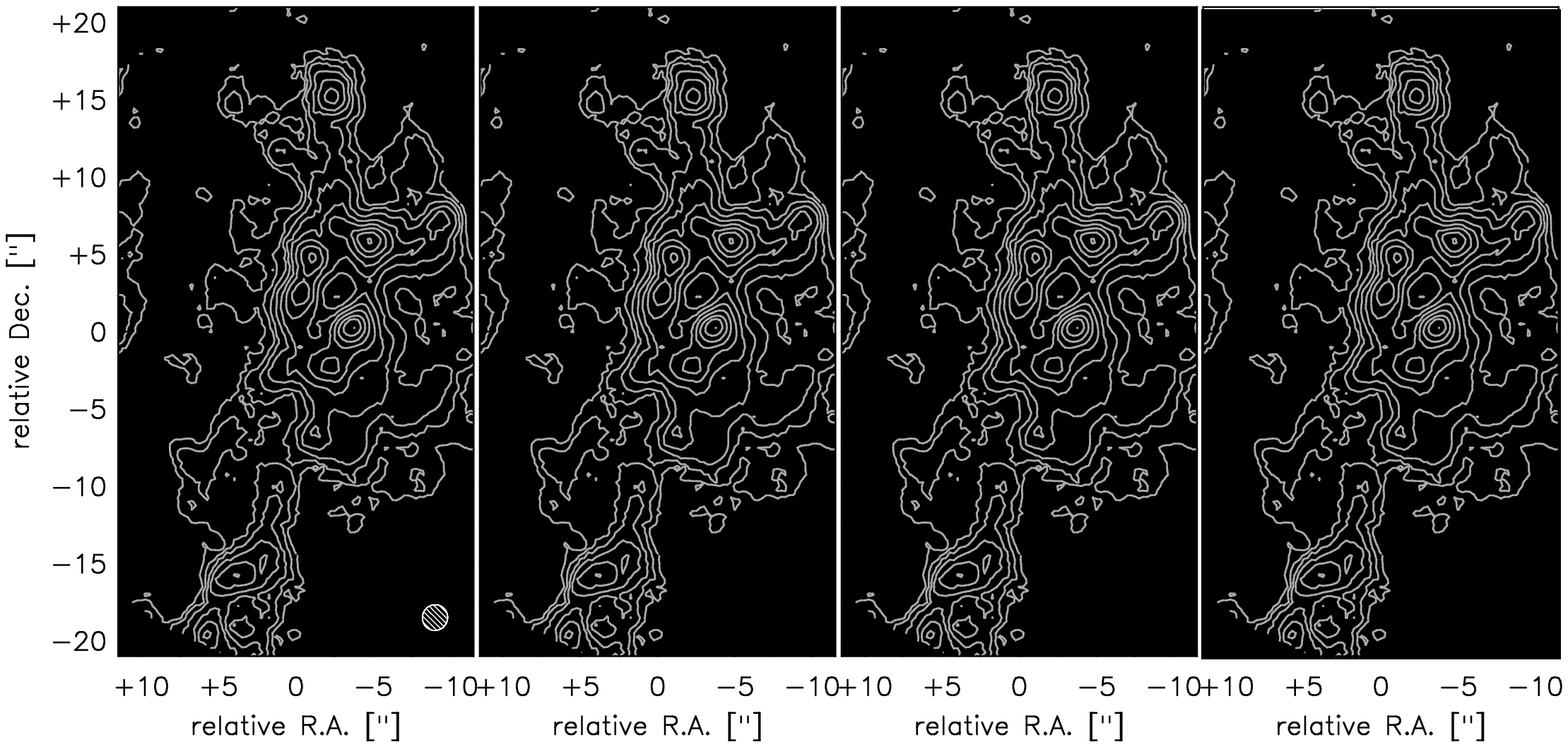}
\caption{Comparison of the \cotwo\ line emission (same contours as
  Fig.  \ref{fig:mom_w}) to the stellar light distribution in the HST
  V band ({\it left}) as well as the H$\alpha$ ({\it middle left}) and
  Pa$\alpha$ ({\it middle right}) line emission mainly arising from
  HII regions and the MIPS 24$\mu$m continuum emission tracing warm
  dust heated by young stars ({\it right}).}
\label{fig:hst_w}      
\end{figure}

\begin{figure}
\centering
\includegraphics[height=18cm,angle=0]{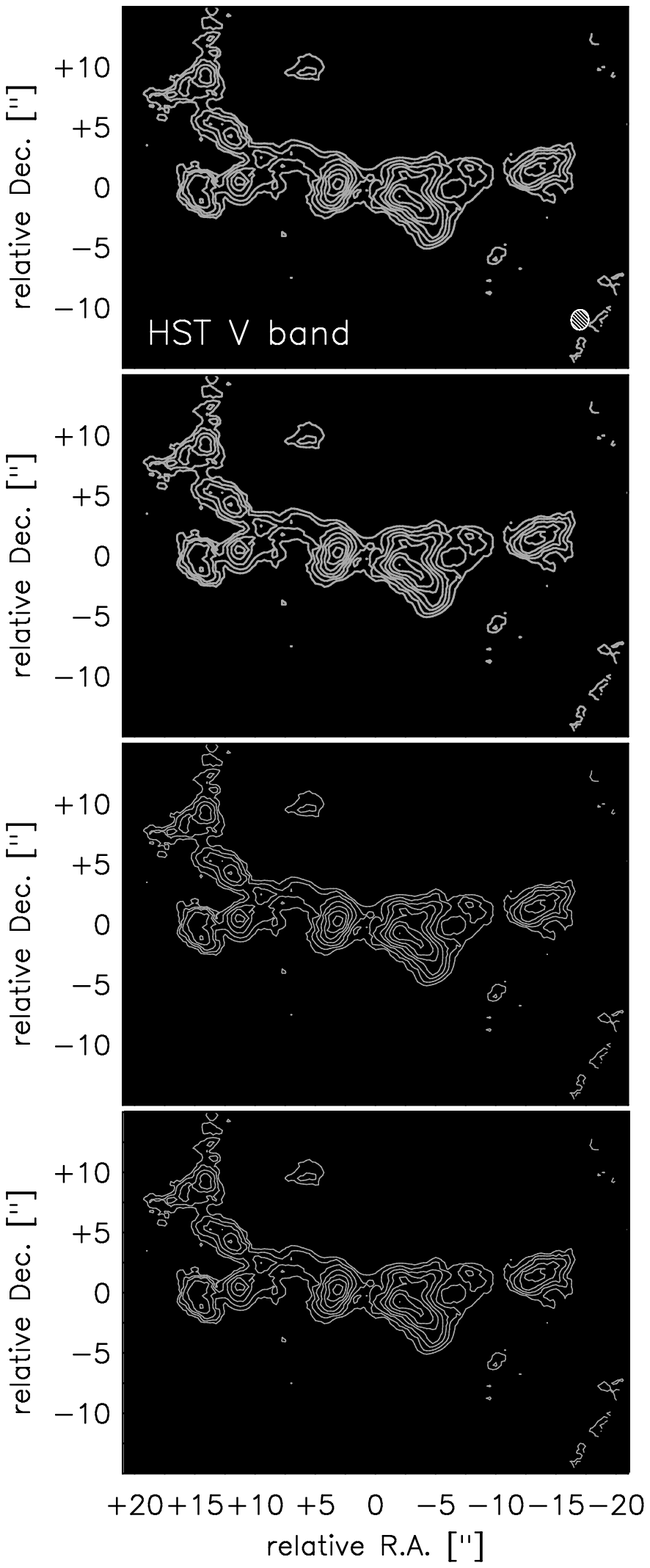}
\caption{Comparison of the \cotwo\ line emission (same contours as
  Fig.  \ref{fig:mom_s}) to the stellar light distribution in the HST
  V band ({\it top}) as well as the H$\alpha$ ({\it middle top}) and
  Pa$\alpha$ ({\it middle bottom}) line emission mainly arising from
  HII regions and the MIPS 24$\mu$m continuum emission tracing warm
  dust heated by young stars ({\it bottom}).}
\label{fig:hst_s}      
\end{figure}

In order to compare the distribution of the molecular gas with the
sites of recent star formation we use our CO data and archival HST
images from \cite{sco01}. As expected the molecular gas coincides very
well with the prominent dust lanes seen in the individual spiral arms.
Using the 1.6'' resolution \cotwo\ data one can see that the emission
peaks do fall into regions of high extinction obvious in the HST V
band image (Fig. \ref{fig:hst_w} and \ref{fig:hst_s}). In particular,
the double gas lane in the brightest part of the western region does
show a line of enhanced V band emission between the two CO lanes.

While enhanced extinction is present across the full width of the gas
spiral arms of $\sim$ 7'' it is not uniform but rather patchy. The
widths of the dust lanes visible in the HST V band image range from
about 0.5'' to 2'' suggesting that the GMCs we resolve in the \cotwo\
data are causing the extinction and that the molecular gas is mainly
distributed within GMCs. An upper limit on the scale height of the gas
disk is about 20--80\,pc if we assume that the extincting material can
not have a larger vertical extent than a dust lane is wide. Combining
these numbers with the derived H$_2$ column densities N(H$_2$) in \S
\ref{sec:lvg} (see Tab. \ref{tab:lvg}) returns approximately the H$_2$
volume densities n(H$_2$) obtained by the LVG analysis. This implies
that most of the GMCs seen at 1.6'' resolution are mostly single
entities and not blends of several GMCs.

Most of the HII regions as traced by their H$\alpha$ and Pa$\alpha$
line emission lie downstream of the molecular gas as it is expected if
star formation is occurring inside the molecular gas spiral arms (see
Fig.  \ref{fig:hst_w} and \ref{fig:hst_s}). Since the southern region
is closer to corotation (where the pattern speed is closer to the
angular velocity) than the western region, this can explain why two of
the \cotwo\ peaks coincide with star formation whereas only one does
so in the western region. Except for one case, all Pa$\alpha$ peaks
have H$\alpha$ emission associated with them suggesting that
extinction is not extreme towards HII regions.

It is interesting to compare the average extinction toward HII regions
derived from the recombination line ratios of H$\alpha$ and Pa$\alpha$
of $\rm A_V \sim 3.0$ \citep{sco01} with those estimated from the CO
data. We use equation (6) from \cite{nak95} of $N(H_2) =
1.87\times10^{21}\,A_V\,[mag] - N(HI) [cm^{-2}]/2$ and our average
$H_2$ column density of $<N(H_2)> = (2.3 - 3.3)\times10^{22}\,\rm
cm^{-2}$. Since the ISM is mostly molecular in the regions studied
here, e.g. \cite{schu07} find values for the ratio of surface density
of $\sum_{HI} / \sum_{H_2} < 0.2$ for r$\le$ 1.6\,kpc and $\sim$ 0.5
for the southern region at 2.6\,kpc, we neglect the contribution from
atomic hydrogen. The derived extinction is about $A_V \sim 12 - 18$,
significantly larger than that measured in the HII regions. This value
is even larger ($A_V \sim (15 - 33)$) then using a Galactic relation
of $N(H_2)/A_V = 10^{21}\,cm^{-2}K^{-1}s$ \citep{boh78}. Thus a lot of
massive star formation could be hidden inside the gas spiral arms and
only be detectable at longer wavelengths such as the infrared and
radio. Using the HiRes deconvolved 24\,$\mu$m MIPS image (Dumas et
al., subm.) and the H$\alpha$ map from \cite{cal05} we tested this
scenario at 2'' resolution. As the 24$\mu$m emission coincides with
the CO emission and is thus mainly arising inside (or upstream) the
H$\alpha$ emission (see Fig. \ref{fig:hst_w} and \ref{fig:hst_s}),
this supports the picture that hidden star formation does occur inside
the spiral arms.  Using the H$\alpha$ emission located in the spiral
arms and 24$\mu$m emission arising from the same area, star formation
rates (SFRs) can be estimated\footnote{We do not subtract the
  contribution from the diffuse emission as star formation appears to
  be the dominant source for the H$\alpha$ and 24$\mu$m emission in
  the spiral arms.}.  Correcting the observed H$\alpha$ flux for the
average extinction of $A_V \sim 3$ found for HII regions \citep{sco01}
and using the SFR prescriptions of \cite{ken98} for H$\alpha$ and of
\cite{rie09} for the 24\,$\mu$m data, we find that the SFR hidden in
the spiral arms is about 50\% of that traced by H$\alpha$. This 'lack'
of (embedded) star formation inside the gas spiral arms is consistent
with the fact that the molecular gas is still cold and might therefore
have not yet started to form new stars.

There is no strong correlation between the location of bright HII
regions and massive CO clouds or holes, though most HII regions tend to
lie next to a \cotwo\ peak. This would be in line with the picture
described by \cite{bas05} who found a correlation between the radius
and mass for clusters of star clusters that is similar to the relation
seen for GMCs. They interpret this as evidence that a GMC does form an
assemble of star clusters. 

\section{The impact of the star formation on the ISM}
\label{sec:dis}

The physical properties of the GMC complexes located in the spiral
arms of M\,51 are very similar to those observed for GMCs in our
Galaxy. Our LVG analysis suggests that the conversion factor between
CO line emission and H$_2$ mass is similar to the standard conversion
factor $X_{CO}$. This result is in apparent contradiction to other
studies conducted in M\,51 at significantly lower angular resolution.
The simplest explanation is that the conversion factor is not constant
across the disk (or even within the spiral arms) but a function of
environment (even for the diffuse gas probed here).
Therefore a spatial resolution matched to the size of GMC complexes is
required to obtain reasonable numbers. Further, the assumption of a
single value is not correct when measuring the total amount of gas
present in a nearby spiral galaxy. Our 65\,pc resolution \cotwo\ data
shows that the spiral arms break up into clumps at these scales that
are roughly aligned along the spiral arms. In the western region the
CO peaks are more abundant toward the leading side of the spiral
showing a slight shift from the peak of the more smooth emission as
traced by, e.g., the 4.5'' resolution \coone\ emission (see Fig.
\ref{fig:w_all} and \ref{fig:mom_w}). No offset is obvious for the
southern region (see Fig. \ref{fig:s_all} and \ref{fig:mom_s}) as
expected as the pattern is moving close to the angular velocity at
these radii.

At our 120\,pc resolution we find no evidence that star formation in
HII regions is impacting the properties of the GMCs that are at about
$\ge$1'' (40\,pc) distance, in agreement with findings in M33 by
\cite{wil97}. Only two positions which are located at the inner edge
of the spiral arm exhibit elevated temperatures that may be related to
the large-scale dynamics.

Based on the observations presented here, we conclude that most of the
GMCs in the spiral arms of M51 are cold, dense structures that are not
heavily affected by the neighboring star forming HII regions. Despite
the inferred high extinction of $\rm A_V \sim (15-30)$ towards these
GMCs only about 50\% of the star formation traced by H$\alpha$
emission is coinciding with the gas spiral arms indicating that most
of the molecular gas seen is not yet actively forming stars.

\section{Summary and Conclusions}
\label{sec:sum}

M\,51 offers the ideal environment to study the physical properties of
the molecular gas in the spiral arms of a disk galaxy. Combining short
spacing corrected interferometric observations of the \coone, \cotwo,
and \cotre\ line emission in two selected arm regions in the central
6\,kpc, a LVG analysis finds that the GMCs inside the spiral are
similar to those observed in our Galaxy. The average kinetic
temperature and H$_2$ density (assuming virialized clouds) are $\rm
T_{kin} = 20\,K$ and $\rm n(H_2) = 120\,cm^{-3}$, respectively, at a
linear resolution of 120\,pc (290\,pc for the southern region).
Similarly, the derived conversion factor $X_{CO}$ is close to the
Galactic value of $\rm 1.8\times10^{20}\,cm^{-2}K^{-1}\,km^{-1}s$
recently derived by \cite{dam01}. We interpret our results such that
the physical properties of the GMC population do not change much as a
function of environment in galaxy disks, consistent with the findings
of \cite{bol08} who analyzed GMC properties in local galaxies.

Comparison of the derived molecular gas properties to the location of
star formation shows no obvious trend suggesting that the massive star
formation has no strong impact onto the GMCs inside the spiral arms.
Comparison of the extinction inferred from the derived $H_2$ density
and measured toward HII regions shows that the extinction toward the
molecular clouds is at least a factor of 5 higher with an average
$A_V$ of 15--30\,mag than the average $A_V$ of 3.1 inferred for HII
regions by \cite{sco01}.

\acknowledgments Based on observations carried out with the IRAM 30m
telescope. IRAM is supported by INSU/CNRS (France), MPG (Germany) and
IGN (Spain). We thank very much K. Schuster and C. Kramer for making
available to us the HERA \cotwo\ data cube and G. Dumas for the HiRes
version of the MIPS 24$\mu$m image as well as her help with the SFRs.
We are thankful for the public access to the 'Nobeyama CO Atlas of
Nearby Spiral Galaxies'. The comments and suggestions by the anonymous
referee were highly appreciated and helped to improve the paper. The
Owens Valley Radio Observatory was funded in part by NSF grant AST
99-81546.

\facility{OVRO, IRAM (30m)}

\bibliography{references}

\begin{thebibliography}{46}
\expandafter\ifx\csname natexlab\endcsname\relax\def\natexlab#1{#1}\fi

\bibitem[{{Aalto} {et~al.}(1999){Aalto}, {H{\"u}ttemeister}, {Scoville}, \&
  {Thaddeus}}]{aal99}
{Aalto}, S., {H{\"u}ttemeister}, S., {Scoville}, N.~Z., \& {Thaddeus}, P. 1999,
  \apj, 522, 165

\bibitem[{{Adler} {et~al.}(1992){Adler}, {Lo}, {Wright}, {Rydbeck}, {Plante},
  \& {Allen}}]{adl92}
{Adler}, D.~S., {Lo}, K.~Y., {Wright}, M.~C.~H., {Rydbeck}, G., {Plante},
  R.~L., \& {Allen}, R.~J. 1992, \apj, 392, 497

\bibitem[{{Bastian} {et~al.}(2005){Bastian}, {Gieles}, {Efremov}, \&
  {Lamers}}]{bas05}
{Bastian}, N., {Gieles}, M., {Efremov}, Y.~N., \& {Lamers}, H.~J.~G.~L.~M.
  2005, \aap, 443, 79

\bibitem[{{Bell} {et~al.}(2007){Bell}, {Whyatt}, {Viti}, \& {Redman}}]{bel07}
{Bell}, T.~A., {Whyatt}, W., {Viti}, S., \& {Redman}, M.~P. 2007, \mnras, 382,
  1139

\bibitem[{{Benford} \& {Staguhn}(2008)}]{ben08}
{Benford}, D.~J., \& {Staguhn}, J.~G. 2008, in Astronomical Society of the
  Pacific Conference Series, Vol. 381, Infrared Diagnostics of Galaxy
  Evolution, ed. {R.-R.~Chary, H.~I.~Teplitz, \& K.~Sheth}, 132--+

\bibitem[{{Bohlin} {et~al.}(1978){Bohlin}, {Savage}, \& {Drake}}]{boh78}
{Bohlin}, R.~C., {Savage}, B.~D., \& {Drake}, J.~F. 1978, \apj, 224, 132

\bibitem[{{Bolatto} {et~al.}(2008){Bolatto}, {Leroy}, {Rosolowsky}, {Walter},
  \& {Blitz}}]{bol08}
{Bolatto}, A.~D., {Leroy}, A.~K., {Rosolowsky}, E., {Walter}, F., \& {Blitz},
  L. 2008, \apj, 686, 948

\bibitem[{{Calzetti} {et~al.}(2005){Calzetti}, {Kennicutt}, {Bianchi},
  {Thilker}, {Dale}, {Engelbracht}, {Leitherer}, {Meyer}, {Sosey}, {Mutchler},
  {Regan}, {Thornley}, {Armus}, {Bendo}, {Boissier}, {Boselli}, {Draine},
  {Gordon}, {Helou}, {Hollenbach}, {Kewley}, {Madore}, {Martin}, {Murphy},
  {Rieke}, {Rieke}, {Roussel}, {Sheth}, {Smith}, {Walter}, {White}, {Yi},
  {Scoville}, {Polletta}, \& {Lindler}}]{cal05}
{Calzetti}, D., {Kennicutt}, Jr., R.~C., {Bianchi}, L., {Thilker}, D.~A.,
  {Dale}, D.~A., {Engelbracht}, C.~W., {Leitherer}, C., {Meyer}, M.~J.,
  {Sosey}, M.~L., {Mutchler}, M., {Regan}, M.~W., {Thornley}, M.~D., {Armus},
  L., {Bendo}, G.~J., {Boissier}, S., {Boselli}, A., {Draine}, B.~T., {Gordon},
  K.~D., {Helou}, G., {Hollenbach}, D.~J., {Kewley}, L., {Madore}, B.~F.,
  {Martin}, D.~C., {Murphy}, E.~J., {Rieke}, G.~H., {Rieke}, M.~J., {Roussel},
  H., {Sheth}, K., {Smith}, J.~D., {Walter}, F., {White}, B.~A., {Yi}, S.,
  {Scoville}, N.~Z., {Polletta}, M., \& {Lindler}, D. 2005, \apj, 633, 871

\bibitem[{{Corder} {et~al.}(2008){Corder}, {Sheth}, {Scoville}, {Koda},
  {Vogel}, \& {Ostriker}}]{cor08}
{Corder}, S., {Sheth}, K., {Scoville}, N.~Z., {Koda}, J., {Vogel}, S.~N., \&
  {Ostriker}, E. 2008, \apj, 689, 148

\bibitem[{{Dame} {et~al.}(2001){Dame}, {Hartmann}, \& {Thaddeus}}]{dam01}
{Dame}, T.~M., {Hartmann}, D., \& {Thaddeus}, P. 2001, \apj, 547, 792

\bibitem[{{Feldmeier} {et~al.}(1997){Feldmeier}, {Ciardullo}, \&
  {Jacoby}}]{fel97}
{Feldmeier}, J.~J., {Ciardullo}, R., \& {Jacoby}, G.~H. 1997, \apj, 479, 231

\bibitem[{{Garcia-Burillo} {et~al.}(1993){Garcia-Burillo}, {Guelin}, \&
  {Cernicharo}}]{gar93}
{Garcia-Burillo}, S., {Guelin}, M., \& {Cernicharo}, J. 1993, \aap, 274, 123

\bibitem[{{Goldreich} \& {Kwan}(1974)}]{gol74}
{Goldreich}, P., \& {Kwan}, J. 1974, \apj, 189, 441

\bibitem[{{Goldsmith}(2001)}]{gol01}
{Goldsmith}, P.~F. 2001, \apj, 557, 736

\bibitem[{{Guelin} {et~al.}(1995){Guelin}, {Zylka}, {Mezger}, {Haslam}, \&
  {Kreysa}}]{gue95}
{Guelin}, M., {Zylka}, R., {Mezger}, P.~G., {Haslam}, C.~G.~T., \& {Kreysa}, E.
  1995, \aap, 298, L29+

\bibitem[{{Kennicutt}(1998)}]{ken98}
{Kennicutt}, Jr., R.~C. 1998, \araa, 36, 189

\bibitem[{{Koda} {et~al.}(2009){Koda}, {Scoville}, {Sawada}, {La Vigne},
  {Vogel}, {Potts}, {Carpenter}, {Corder}, {Wright}, {White}, {Zauderer},
  {Patience}, {Sargent}, {Bock}, {Hawkins}, {Hodges}, {Kemball}, {Lamb},
  {Plambeck}, {Pound}, {Scott}, {Teuben}, \& {Woody}}]{kod09}
{Koda}, J., {Scoville}, N., {Sawada}, T., {La Vigne}, M.~A., {Vogel}, S.~N.,
  {Potts}, A.~E., {Carpenter}, J.~M., {Corder}, S.~A., {Wright}, M.~C.~H.,
  {White}, S.~M., {Zauderer}, B.~A., {Patience}, J., {Sargent}, A.~I., {Bock},
  D.~C.~J., {Hawkins}, D., {Hodges}, M., {Kemball}, A., {Lamb}, J.~W.,
  {Plambeck}, R.~L., {Pound}, M.~W., {Scott}, S.~L., {Teuben}, P., \& {Woody},
  D.~P. 2009, \apjl, 700, L132

\bibitem[{{Kramer} {et~al.}(2005){Kramer}, {Mookerjea}, {Bayet},
  {Garcia-Burillo}, {Gerin}, {Israel}, {Stutzki}, \& {Wouterloot}}]{kra05}
{Kramer}, C., {Mookerjea}, B., {Bayet}, E., {Garcia-Burillo}, S., {Gerin}, M.,
  {Israel}, F.~P., {Stutzki}, J., \& {Wouterloot}, J.~G.~A. 2005, \aap, 441,
  961

\bibitem[{{Langer} \& {Penzias}(1993)}]{lan93}
{Langer}, W.~D., \& {Penzias}, A.~A. 1993, \apj, 408, 539

\bibitem[{{Lee} {et~al.}(2005){Lee}, {Chandar}, \& {Whitmore}}]{lee05}
{Lee}, M.~G., {Chandar}, R., \& {Whitmore}, B.~C. 2005, \aj, 130, 2128

\bibitem[{{Meier} \& {Turner}(2001)}]{mei01}
{Meier}, D.~S., \& {Turner}, J.~L. 2001, \apj, 551, 687

\bibitem[{{Meier} {et~al.}(2008){Meier}, {Turner}, \& {Hurt}}]{mei08}
{Meier}, D.~S., {Turner}, J.~L., \& {Hurt}, R.~L. 2008, \apj, 675, 281

\bibitem[{{Meijerink} {et~al.}(2005){Meijerink}, {Tilanus}, {Dullemond},
  {Israel}, \& {van der Werf}}]{meij05}
{Meijerink}, R., {Tilanus}, R.~P.~J., {Dullemond}, C.~P., {Israel}, F.~P., \&
  {van der Werf}, P.~P. 2005, \aap, 430, 427

\bibitem[{{Milam} {et~al.}(2005){Milam}, {Savage}, {Brewster}, {Ziurys}, \&
  {Wyckoff}}]{mil05}
{Milam}, S.~N., {Savage}, C., {Brewster}, M.~A., {Ziurys}, L.~M., \& {Wyckoff},
  S. 2005, \apj, 634, 1126

\bibitem[{{Minamidani} {et~al.}(2008){Minamidani}, {Mizuno}, {Mizuno},
  {Kawamura}, {Onishi}, {Hasegawa}, {Tatematsu}, {Ikeda}, {Moriguchi},
  {Yamaguchi}, {Ott}, {Wong}, {Muller}, {Pineda}, {Hughes}, {Staveley-Smith},
  {Klein}, {Mizuno}, {Nikoli{\'c}}, {Booth}, {Heikkil{\"a}}, {Nyman}, {Lerner},
  {Garay}, {Kim}, {Fujishita}, {Kawase}, {Rubio}, \& {Fukui}}]{min08}
{Minamidani}, T., {Mizuno}, N., {Mizuno}, Y., {Kawamura}, A., {Onishi}, T.,
  {Hasegawa}, T., {Tatematsu}, K., {Ikeda}, M., {Moriguchi}, Y., {Yamaguchi},
  N., {Ott}, J., {Wong}, T., {Muller}, E., {Pineda}, J.~L., {Hughes}, A.,
  {Staveley-Smith}, L., {Klein}, U., {Mizuno}, A., {Nikoli{\'c}}, S., {Booth},
  R.~S., {Heikkil{\"a}}, A., {Nyman}, L.-{\AA}., {Lerner}, M., {Garay}, G.,
  {Kim}, S., {Fujishita}, M., {Kawase}, T., {Rubio}, M., \& {Fukui}, Y. 2008,
  \apjs, 175, 485

\bibitem[{{Nakai} \& {Kuno}(1995)}]{nak95}
{Nakai}, N., \& {Kuno}, N. 1995, \pasj, 47, 761

\bibitem[{{Nakai} {et~al.}(1994){Nakai}, {Kuno}, {Handa}, \& {Sofue}}]{nak94}
{Nakai}, N., {Kuno}, N., {Handa}, T., \& {Sofue}, Y. 1994, \pasj, 46, 527

\bibitem[{{Rand} \& {Kulkarni}(1990)}]{ran90}
{Rand}, R.~J., \& {Kulkarni}, S.~R. 1990, \apjl, 349, L43

\bibitem[{{Rieke} {et~al.}(2009){Rieke}, {Alonso-Herrero}, {Weiner},
  {P{\'e}rez-Gonz{\'a}lez}, {Blaylock}, {Donley}, \& {Marcillac}}]{rie09}
{Rieke}, G.~H., {Alonso-Herrero}, A., {Weiner}, B.~J.,
  {P{\'e}rez-Gonz{\'a}lez}, P.~G., {Blaylock}, M., {Donley}, J.~L., \&
  {Marcillac}, D. 2009, \apj, 692, 556

\bibitem[{{Roberts} \& {Stewart}(1987)}]{rob87}
{Roberts}, Jr., W.~W., \& {Stewart}, G.~R. 1987, \apj, 314, 10

\bibitem[{{Sanders} {et~al.}(1984){Sanders}, {Solomon}, \& {Scoville}}]{san84}
{Sanders}, D.~B., {Solomon}, P.~M., \& {Scoville}, N.~Z. 1984, \apj, 276, 182

\bibitem[{{Sault} {et~al.}(1995){Sault}, {Teuben}, \& {Wright}}]{sau95}
{Sault}, R.~J., {Teuben}, P.~J., \& {Wright}, M.~C.~H. 1995, in Astronomical
  Society of the Pacific Conference Series, Vol.~77, Astronomical Data Analysis
  Software and Systems IV, ed. R.~A. {Shaw}, H.~E. {Payne}, \& J.~J.~E.
  {Hayes}, 433--+

\bibitem[{{Scheepmaker} {et~al.}(2009){Scheepmaker}, {Lamers}, {Anders}, \&
  {Larsen}}]{sche09}
{Scheepmaker}, R.~A., {Lamers}, H.~J.~G.~L.~M., {Anders}, P., \& {Larsen},
  S.~S. 2009, \aap, 494, 81

\bibitem[{{Schuster} {et~al.}(2007){Schuster}, {Kramer}, {Hitschfeld},
  {Garcia-Burillo}, \& {Mookerjea}}]{schu07}
{Schuster}, K.~F., {Kramer}, C., {Hitschfeld}, M., {Garcia-Burillo}, S., \&
  {Mookerjea}, B. 2007, \aap, 461, 143

\bibitem[{{Scoville} {et~al.}(1994){Scoville}, {Carlstrom}, {Padin}, {Sargent},
  {Scott}, \& {Woody}}]{sco94}
{Scoville}, N., {Carlstrom}, J., {Padin}, S., {Sargent}, A., {Scott}, S., \&
  {Woody}, D. 1994, in Astronomical Society of the Pacific Conference Series,
  Vol.~59, IAU Colloq. 140: Astronomy with Millimeter and Submillimeter Wave
  Interferometry, ed. M.~{Ishiguro} \& J.~{Welch}, 10--+

\bibitem[{{Scoville} \& {Young}(1983)}]{sco83}
{Scoville}, N., \& {Young}, J.~S. 1983, \apj, 265, 148

\bibitem[{{Scoville} {et~al.}(2001){Scoville}, {Polletta}, {Ewald}, {Stolovy},
  {Thompson}, \& {Rieke}}]{sco01}
{Scoville}, N.~Z., {Polletta}, M., {Ewald}, S., {Stolovy}, S.~R., {Thompson},
  R., \& {Rieke}, M. 2001, \aj, 122, 3017

\bibitem[{{Scoville} \& {Sanders}(1987)}]{sco87}
{Scoville}, N.~Z., \& {Sanders}, D.~B. 1987, in Astrophysics and Space Science
  Library, Vol. 134, Interstellar Processes, ed. {D.~J.~Hollenbach \&
  H.~A.~Thronson Jr.}, 21--50

\bibitem[{{Scoville} \& {Solomon}(1974)}]{sco74}
{Scoville}, N.~Z., \& {Solomon}, P.~M. 1974, \apjl, 187, L67+

\bibitem[{{Shetty} {et~al.}(2007){Shetty}, {Vogel}, {Ostriker}, \&
  {Teuben}}]{shet07}
{Shetty}, R., {Vogel}, S.~N., {Ostriker}, E.~C., \& {Teuben}, P.~J. 2007, \apj,
  665, 1138

\bibitem[{{Solomon} {et~al.}(1979){Solomon}, {Sanders}, \& {Scoville}}]{sol79}
{Solomon}, P.~M., {Sanders}, D.~B., \& {Scoville}, N.~Z. 1979, \apjl, 232, L89

\bibitem[{{Tielens}(2005)}]{tie05}
{Tielens}, A.~G.~G.~M. 2005, {The Physics and Chemistry of the Interstellar
  Medium}, ed. A.~G.~G.~M. {Tielens}

\bibitem[{{Tosaki} {et~al.}(2002){Tosaki}, {Hasegawa}, {Shioya}, {Kuno}, \&
  {Matsushita}}]{tos02}
{Tosaki}, T., {Hasegawa}, T., {Shioya}, Y., {Kuno}, N., \& {Matsushita}, S.
  2002, \pasj, 54, 209

\bibitem[{{van der Hulst} {et~al.}(1988){van der Hulst}, {Kennicutt}, {Crane},
  \& {Rots}}]{vdhul88}
{van der Hulst}, J.~M., {Kennicutt}, R.~C., {Crane}, P.~C., \& {Rots}, A.~H.
  1988, \aap, 195, 38

\bibitem[{{Wei{\ss}} {et~al.}(2001){Wei{\ss}}, {Neininger}, {H{\"u}ttemeister},
  \& {Klein}}]{wei01}
{Wei{\ss}}, A., {Neininger}, N., {H{\"u}ttemeister}, S., \& {Klein}, U. 2001,
  \aap, 365, 571

\bibitem[{{Wilson} {et~al.}(1997){Wilson}, {Walker}, \& {Thornley}}]{wil97}
{Wilson}, C.~D., {Walker}, C.~E., \& {Thornley}, M.~D. 1997, \apj, 483, 210

\end{thebibliography}

\clearpage

%%%%% Fig. 1  %%%%
%%%%

\clearpage
%%%%% Fig. 2  %%%%
%%%%

\clearpage

%%%%% Fig. 3  %%%%
%%%%

\clearpage

%%%%% Fig. 4  %%%%
%%%%

\clearpage

%%%%% Fig. 5  %%%%
%%%%

\clearpage

%%%%% Fig. 6  %%%%
%%%%

\clearpage

%%%%% Fig. 7  %%%%
%%%%

\clearpage

%%%%% Fig. 8  %%%%
%%%%

\clearpage

%%%%% Fig. 9  %%%%
%%%%

\clearpage

%%%%% Fig. 10  %%%%
%%%%

\clearpage

%%%%% Fig. 11  %%%%
%%%%

\clearpage

%%%%% Fig. 12  %%%%
%%%%

\clearpage

%%%% Table 1
\begin{deluxetable}{lrr}
\tablecaption{M\,51a fields\label{tab:poi}}
\tablewidth{0pt}
\tablehead{
\colhead{Pointing} & 
\colhead{R.A. (J2000)} & 
\colhead{Dec. (J2000)}
}
\startdata
West  & 13:29:50.14  &  +47:11:25.27 \\
South & 13:29:55.17  &  +47:10:47.41 \\
\enddata
\tablecomments{Pointing centers for two selected fields in the M1
  spiral arm of M\,51a.}
\end{deluxetable}

%%%% Table 2
\begin{deluxetable}{llrccc}
\tablecaption{Observations and Set-up\label{tab:spec}}
\tablewidth{0pt}
\tablehead{
\colhead{Pointing} &
\colhead{Transition} & 
\colhead{Tracks} &
\colhead{$v_{LSR}$} &
\colhead{$\Delta\,v$} & 
\colhead{\# of Channels}\\
&&&
\colhead{($\kms$)} &
\colhead{($\kms$)} &
}
\startdata
West  & \cotwo    & 1.5L,1E,1H & 489 &  2.6 & 64 \\
      & \cotre    & 1C,1L,1.5H & 489 &  2.7 & 32 \\
      & \coeig    & 1C,1L,1.5H & 488 &  5.5 & 16 \\
      & \hcn      & 1C,2L,1.5H & 493 &  6.8 & 16 \\
      & \hco      & 1C,2L,1.5H & 493 &  6.7 & 16 \\
\tableline
South & \cotwo    & 1.5L,1E,1H & 517 &  5.2 & 32 \\
      & \cotre    & 1C,1L,1H   & 517 &  5.4 & 16 \\
      & \coeig    & 1C,1L,1H   & 517 &  5.5 & 16 \\
      & \hcn      & 0.5C,4L,2H & 523 &  6.8 & 16 \\
      & \hco      & 0.5C,4L,2H & 523 &  6.7 & 16 \\ 
\enddata
\tablecomments{Here we list the final spectral resolution used and the
$v_{LSR}$ of the central channel. The observed tracks are given as the
number of full ($\sim$ 5\,hr on-source) tracks per configuration, e.g.
1.5L = 1.5 full tracks in L configuration.}
\end{deluxetable}

%%%% Table 3
\begin{deluxetable}{llccc}
\tablecaption{Parameters of the OVRO Data\label{tab:data}}
\tablewidth{0pt}
\tablehead{
\colhead{Pointing} &
\colhead{Line} & 
\colhead{Weighting}&
\colhead{rms} & 
\colhead{beam} \\
&& &
\colhead{(mJy\,beam$^{-1}$)} & 
\colhead{($\as \times \as$)} 
}
\startdata
West  & \cotwo    & na &  22 & 1.67 $\times$ 1.63 \\
      & \cotwo    & ro &  26 & 1.31 $\times$ 1.28 \\
      & \cotwo    & un &  27 & 1.27 $\times$ 1.25 \\
      & \cotre    & na &  17 & 3.91 $\times$ 3.19 \\
      & \cotre    & ro &  19 & 2.87 $\times$ 2.45 \\   
      & \coeig    & na &  13 & 4.22 $\times$ 3.77 \\
      & \hcn      & na &  13 & 4.76 $\times$ 3.98 \\
      & \hco      & na &  12 & 4.67 $\times$ 3.93 \\
\tableline                          
South & \cotwo    & na &  19 & 1.68 $\times$ 1.46 \\
      & \cotwo    & ro &  22 & 1.33 $\times$ 1.21 \\
      & \cotwo    & un &  25 & 1.28 $\times$ 1.20 \\
      & \cotre    & na &  17 & 4.45 $\times$ 3.69 \\
      & \coeig    & na &  15 & 4.63 $\times$ 4.18 \\
      & \hcn      & na &  12 & 3.88 $\times$ 3.28 \\
      & \hco      & na &  11 & 4.07 $\times$ 3.53 \\
\enddata
\tablecomments{Three different values are used for 
weighting of the $uv$ data, namely a (Briggs) robust parameter of 
+5 (=na; 'natural weighting'), 0 (=ro; 'robust weighting'), and -5 
(=un; 'uniform weighting').}
\end{deluxetable}

%%%% Table 4
\begin{deluxetable}{lcccc}
\tablecaption{Sensitivity of the Short Spacing Corrected Data\label{tab:rms}}
\tablewidth{0pt}
\tablehead{
\colhead{Pointing} &
\colhead{\coone} & 
\colhead{\cotwo}&
\colhead{\cotre} & 
\colhead{\coeig} \\
&
\colhead{(K)} & 
\colhead{(K)} & 
\colhead{(K)} & 
\colhead{(K)} 
}
\startdata
West-2.9  & 0.32 & 0.13 & 0.28 & ---  \\
West-4.5  & 0.23 & 0.08 & 0.09 & 0.08 \\
South-4.5 & 0.25 & 0.06 & 0.09 & 0.09 \\
\enddata
\tablecomments{The {\sl rms} noise in the short spacing corrected data
cubes of the CO transitions. All CO data cubes have the same
resolution of 2.9$\as$ (West-2.9) or 4.5$\as$ (West-4.5, South-4.5).}
\end{deluxetable}

\clearpage

%%%% Table 5
\begin{deluxetable}{lcrrrrrr}
%\rotate
\tablecaption{Line Fluxes and Ratios at 2.9'' resolution\label{tab:flux}}
\tablewidth{0pt}
\tablehead{
\colhead{Position} &
\colhead{Offset} &
\colhead{\coone} & 
\colhead{\cotwo}&
\colhead{\cotre} & 
\colhead{$v_{50}$}&
\colhead{$\frac{^{12}CO(2-1)}{^{12}CO(1-0)}$}&
\colhead{$\frac{^{12}CO(1-0)}{^{13}CO(1-0)}$}
\\
&
\colhead{('';'')} &
\colhead{(K\,$\kms$)} & 
\colhead{(K\,$\kms$)} & 
\colhead{(K\,$\kms$)} & 
\colhead{($\kms$)}& & }
\startdata
w1 & 0;+15& 187 & 116 & 26 &  60 & 0.72 & 5.5\\
w2 & 0;+11& 150 &  87 & 21 &  50 & 0.60 & 5.5\\
w3 & -6;-6&  87 & 103 & $\le$19  & 36 & 1.10 & $\ge$9.5\\
w4 & -2;+6& 193 & 152 & 39 &  35 & 0.65 & 4.3\\
w5 & +1;+5& 219 & 135 & 14 &  60 & 0.70 & 7.5\\
w6 & -1;+3& 204 & 112 & 22 &  50 & 0.60 & 7.0\\
w7 & +2;+2& 185 & 137 & 21 &  46 & 0.71 & 4.5\\
w8 & -1;0 & 303 & 157 & 31 &  40 & 0.55 & 7.5\\
w9 & -2;-3& 222 & 117 & $\le$23  & 45 & 0.55 & $\ge$9.0\\
w10& +5;-4& 102 &  65 & $\le$22  & 43 & 0.63 & $\ge$7.0\\
w11& +1;-5& 185 & 102 & 31 &  30 & 0.45 & 7.7\\
w12& +3;-9& 128 &  64 & 23 &  31 & 0.45 & 7.3\\
w13&+4;-12& 142 &  71 & 28 &  36 & 0.47 & 6.5\\
w14&+6;-15& 215 & 113 & 47 &  38 & 0.42 & 4.5\\
\enddata 

\tablecomments{Here we list the measured integrated line fluxes for
  the different position in the western pointing. The lower case
  letters refer to the 2.9$\as$ resolution data that was used to
  derive the measurements. The $x-$ and $y-$offsets are relative to
  the pointing centers given in Tab. \ref{tab:poi}. The FWHM
  ($v_{50}$) is derived from the \coone\ line. The line ratios
  were derived from scaling the \cotwo, \cotre, and \coeig\ spectra to
  the \coone\ spectrum. This method is more reliable than using the
  peak flux or the integrated flux in the case of low S/N data.
  Therefore the line ratios listed are not in perfect agreement with
  the ratio of the integrated line fluxes.  }
\end{deluxetable}

%%%%%%%%%%%
%%% table 6
\begin{deluxetable}{lcrrrrrrrr}
\rotate
\tablecaption{Line Fluxes and Ratios at 4.5'' resolution\label{tab:fluxa}}
\tablewidth{0pt}
\tablehead{
\colhead{Position} &
\colhead{Offset} &
\colhead{\coone} & 
\colhead{\cotwo}&
\colhead{\cotre} & 
\colhead{\coeig} &
\colhead{$v_{50}$}&
\colhead{$\frac{^{12}CO(2-1)}{^{12}CO(1-0)}$}&
\colhead{$\frac{^{12}CO(1-0)}{^{13}CO(1-0)}$}&
\colhead{$\frac{^{12}CO(1-0)}{C^{18}O(1-0)}$}
\\
&
\colhead{('';'')} &
\colhead{(K\,$\kms$)} & 
\colhead{(K\,$\kms$)} & 
\colhead{(K\,$\kms$)} & 
\colhead{(K\,$\kms$)} &
\colhead{($\kms$)}& & & }
\startdata
W1 & 0;+15& 164 &  91 & 22 & $\le$13 & 64 & 0.63 & 5.1 & $\ge$15 \\
W2 & 0;+11& 151 &  77 & 19 & 10 & 55 & 0.60 & 5.8 & 17  \\
W3 & -6;+7&  83 &  89 &  4 & $\le$9 & 45 & 1.10 & 10& $\ge$10\\
W4 & -2;+6& 184 & 127 & 25 &  7 & 39 & 0.60 & 6.0 & 25 \\
W5 & +2;+4& 155 & 115 & 18 & $\le$11 & 55 & 0.76 & 5.7 & $\ge$14 \\
W6 & -1;0&  273 & 136 & 27 &  6 & 42 & 0.53 & 8.0 & 25 \\
W7 & +1;-5& 170 & 100 & 18 & $\le$7 & 32 & 0.50 & 9.0 & $\ge$26 \\
W8 & +5;-10& 113 &  67 & 16 & $\le$8 & 39 & 0.54 & 7.8 & $\ge$14 \\
W9 & +7;-16& 178 & 100 & 25 & $\le$8 & 39 & 0.45 & 5.0 & $\ge$22 \\
\tableline
S1 & -14;+2& 44 &  46 & $\le$5 & $\le$5 & 26 & 0.87 & $\ge$8.6 & $\ge$9 \\
S2 & -5;-3& 147 &  57 & 11 & $\le$6 & 34 & 0.42 & 11 & $\ge$23\\
S3 & -2;+1& 134 &  62 &  7 & $\le$7 & 37 & 0.48 & 9.0 & $\ge$19\\
S4 & +3;+1& 166 &  59 &  5 & $\le$9 & 46 & 0.43 & 11 & $\ge$19\\
S5A& +5;-1&  45 &  26 & $\le$6 & $\le$6 & 30 & 0.52 & $\ge$7.5 & $\ge$8\\
S5B&      &  86 &  29 & $\le$6 & $\le$6 & 30 & 0.37 & $\ge$9.5 & $\ge$15\\
S6A& +13;0&  20 &  24 & $\le$6 & $\le$8 & 40 & 0.90 & 6.5 & $\ge$3\\
S6B&       & 80 &  36 &  8 & $\le$4 & 22 & 0.40 & 7.3  & $\ge$20\\
S7A& +12;+5& 60 &  26 &  3 & $\le$6 & 33 & 0.55 & 10 & $\ge$10\\
S7B&       & 32 &  21 &  2 & $\le$5 & 26 & 0.71 & 14  & $\ge$6\\
\enddata 
\tablecomments{Here we list the measured integrated line fluxes for
  the different position in the western and southern pointing. The
  capital letters refer to the 4.5$\as$ resolution data where all
 measurements were obtained. See
  caption of Tab. \ref{tab:flux} for details. }
\end{deluxetable}

%%%% Table 7
\begin{deluxetable}{lc|rl|rl|rl|rl}
\rotate
\tablecaption{Results of the Large Velocity Gradient Analysis\label{tab:lvg}}
\tablewidth{0pt}
\tablehead{
\colhead{Position} &
\colhead{$\frac{\rm d\it{v}}{\rm d\it{r}}$}&
\multicolumn{2}{c}{$\rm T_{kin}$} & 
\multicolumn{2}{c}{$\rm N(H_2)$}&
\multicolumn{2}{c}{$\rm n(H_2)$} &
\multicolumn{2}{c}{$X_{CO}$} 
\\
\colhead{} &
\colhead{} &
\multicolumn{2}{c}{(K)} & 
\multicolumn{2}{c}{($\rm 10^{22}cm^{-2}$)} & 
\multicolumn{2}{c}{($\rm cm^{-3}$)} &
\multicolumn{2}{c}{($\rm 10^{20}cm^{-2}K^{-1}km^{-1}s$)}
\\
\colhead{} &
\colhead{} &
\colhead{$\frac{\rm d\it{v}}{\rm d\it{r}}$=c.} &
\colhead{vir.} &
\colhead{$\frac{\rm d\it{v}}{\rm d\it{r}}$=c.} &
\colhead{vir.} &
\colhead{$\frac{\rm d\it{v}}{\rm d\it{r}}$=c.} &
\colhead{vir.} &
\colhead{$\frac{\rm d\it{v}}{\rm d\it{r}}$=c.} &
\colhead{vir.} 
}
\startdata
w1   & 0.4  &  22  &  27  &  1.5 &   1.7 &  340  &  210 &  0.8 &   0.9 \\
w2   & 0.4  &  13  &  13  &  1.8 &   2.1 &  280  &  160 &  1.2 &   1.4 \\
w4\tablenotemark{\ast}   & 0.5  &  12  &  13  &  2.8 &   3.5 &  350  &  230 &  1.5 &   1.8 \\
w5   & 0.3  &  35  &  50  &  1.2 &   1.2 &  250  &  120 &  0.6 &   0.5 \\
w6   & 0.3  &  18  &  22  &  1.6 &   2.4 &  230  &  120 &  0.8 &   1.2 \\
w7   & 0.5  &  16  &  18  &  2.1 &   2.6 &  380  &  250 &  1.1 &   1.4 \\
w8   & 0.3  &  14  &  16\tablenotemark{\dag}  &  2.9 &   4.5 &  200  &  100 &  0.9 &   1.5 \\
w11  & 0.3  &   8  &  10\tablenotemark{\dag}  &  3.2 &   4.3 &  180  &   90 &  1.7 &   2.3 \\
w12  & 0.3  &   8  &  10  &  2.2 &   2.9 &  190  &  100 &  1.7 &   2.3 \\
w14  & 0.4  &  5\tablenotemark{\dag} &  5\tablenotemark{\dag}  & 10.9 &  16.6 &  280  &  170 &  5.1 &   7.7 \\
\tableline
S2   & 0.2  &  13  & 15\tablenotemark{\dag}  &  1.4 &   2.6 &  120  &   50 &  0.9 &   1.8 \\
S3\tablenotemark{\ast}   & 0.3  &  13  &  17  &  1.3 &   2.0 &  160  &   70 &  1.0 &   1.5 \\
S4\tablenotemark{\ast}   & 0.2  &  13  &  15  &  1.5 &   2.9 &  120  &   50 &  0.9 &   1.8 \\
S6B\tablenotemark{\ast}  & 0.3  &  7\tablenotemark{\dag}  &  7\tablenotemark{\dag}  &  1.7 &   3.0 &  190  &  100 &  2.1 &   3.7 \\
S7A  & 0.2  &  21  &  24  &  0.4 &   0.7 &  140  &   60 &  0.7 &   1.2 \\
S7B\tablenotemark{a}  & 0.3  &  43  &  50  &  0.2 &   0.3 &  260  &  110 &  0.5 &   0.8 \\
\tableline
$< \rm value >$     & 0.3  &  16  &  20  &  2.3 &   3.3 &  240  &  120 &  1.3 &   2.0 \\
\enddata

\tablenotetext{a}{A $\rm [^{12}CO]/[^{13}CO]$ abundance of 60.0 was used.}
\tablenotetext{\ast}{Bright H$\alpha$ emission present inside CO
  aperture.}  
\tablenotetext{\dag}{Beam filling factor $>$ 1.}

\tablecomments{The kinetic temperatures $\rm T_{kin}$ and H$_2$ column
  and volume densities N(H$_2$) and n(H$_2$) were derived from the LVG
  analysis described in section \ref{sec:lvg}. Typical uncertainties
  for the kinetic temperature are (75-100)\% while the H$_2$ densities
  have uncertainties of about 25\% based on the (10-15)\% calibration
  uncertainty for the CO line data. A CO abundance of $\rm
  [^{12}CO]/[H_2]$ and $\rm [^{12}CO]/[^{13}CO]$ of $8.0\times10^{-5}$
  and 30.0, respectively, were assumed. The velocity gradient $\rm
  \frac{d\it{v}}{d\it{r}}$ was either kept fixed at 1.0
  $\kms\,pc^{-1}$ or virialized gas was assumed where $\rm
  \frac{d\it{v}}{d\it{r}}=3.1\sqrt{\frac{n(H_2)}{10^4}}$. The
  corresponding conversion factors $X_{CO}$ from CO line temperature
  to H$_2$ column density are also listed, the standard Galactic
  conversion factor is $\rm 1.8\times10^{20}\,cm^{-2}K^{-1}km^{-1}s$
  \citep[e.g.  ][]{dam01}. The last row lists the average values of
  all regions analyzed. }

\end{deluxetable}

\end{document}